\documentclass{aastex63}
\received{July 6, 2020}
\revised{November 12, 2020}
\accepted{November 29, 2020}
\submitjournal{AJ}

\shorttitle{Variability and its impact on [Fe/H] determination}

\shortauthors{Llancaqueo Albornoz et al.}

\begin{document}
\title{Variability in NGC~3201 giant stars and its impact on their spectroscopic [Fe/H] determination}

\correspondingauthor{Álvaro Llancaqueo Albornoz}
\email{alllancaqueo@udec.cl}

\author[0000-0003-3131-2464]{Á. Llancaqueo Albornoz}
\affiliation{Departamento de Astronomía, Casilla 160-C, Universidad de Concepción, Concepción, Chile}

\author[0000-0001-6205-1493]{S. Villanova}
\affiliation{Departamento de Astronomía, Casilla 160-C, Universidad de Concepción, Concepción, Chile}

\author{C.C. Cortés}
\affiliation{Departamento de Astronomía, Casilla 160-C, Universidad de Concepción, Concepción, Chile}
\affiliation{Departamento de Física, Facultad de Ciencias, Universidad del Bío-Bío, Avenida Collao 1202, Casilla 15-C, Concepción, Chile}

\author[0000-0002-7091-5025]{J.A. Ahumada}
\affiliation{Observatorio Astron\'omico, Universidad Nacional de C\'ordoba, Laprida 854, 5000, C\'ordoba, Argentina.}

\author[0000-0001-9382-3986]{C. Parisi}
\affiliation{Observatorio Astron\'omico, Universidad Nacional de C\'ordoba, Laprida 854, 5000, C\'ordoba, Argentina.}
\affiliation{Instituto de Astronom{\'\i}a Te\'orica y Experimental (CONICET-UNC), Laprida 854, 5000, C\'ordoba, Argentina.}

\begin{abstract}
We present the analysis of 34 light curves in $V$ and  $I$ of 17 giant stars in the globular cluster NGC~3201, to check if such stars are variable and if their variability has some kind of impact on the iron abundance as obtained from spectroscopic measurements. First, we computed the Generalized Lomb-Scargle and Phase Dispersion Minimization periodograms on the sample  to check if the stars were variables. In this way, 7 stars of the sample were found to be non-variable, 2 stars are considered as possible variables, and 8 stars were found to be variable, with periods ranging from $0.0881\pm0.0001$ to $0.5418\pm0.0027$ days. According to the literature, the variables have distinct values of $\text{[Fe I/H]}$: the 3 most metal-rich stars are in the RGB stage, one has an $\text{[Fe I/H]}=-1.37$~dex, while the other two have $\text{[Fe I/H]}=-1.31$~dex. The two most metal-poor variables have $\text{[Fe I/H]}=-1.61$~dex and $\text{[Fe I/H]}=-1.62$~dex, and are AGB stars; the remaining variables have $\text{[Fe I/H]}=-1.44$, $-1.48$, and $-1.50$ dex, the first two being RGB while the last is AGB star. On the other hand, stars that appear to be non-variables have $-1.56\leq\text{[Fe I/H]}\leq-1.40$. We conclude that variability somehow affects the spectroscopic determination of the iron content of giant stars in NGC~3201 increasing the iron spread of the cluster. If variability is not taken into account, this spread could be wrongly interpreted as due to an intrinsic iron spread affecting the stars of the cluster.
%250 words
\end{abstract}

\keywords{globular clusters: individual (NGC 3201) --- stars: abundances}

\section{Introduction}\label{sec:intro}
In several respects, Globular Clusters (GCs) have remained as one of the most  relevant astronomical objects of study during more than one hundred years. They were at first thought to be examples of simple stellar populations, which was  later proven incorrect  due to detailed chemical analyzes showing that almost all GCs display inhomogeneities in their light element contents \citep[e.g.,][]{2004ARA&A..42..385G, 2009A&A...505..139C,2009A&A...505..117C}. The most prominent feature is the Na-O anticorrelation, with Ruprecht~106 \citep{2013ApJ...778..186V} being the sole exception to date. But there are also GCs that display a metallicity distribution or iron spread: $\omega$~Centauri \citep[e.g.,][]{2003ApJ...591..916O, 2011A&A...527A..18P, 2011ApJ...731...64M} and Terzan~5 \citep[e.g.,][]{2011ApJ...726L..20O, 2013ApJ...779L...5O}, both displaying a multi-modal distribution spanning a range of $\sim1$~dex, and  M54 \citep[e.g.,][]{2008AJ....136.1147B, 2010A&A...520A..95C, 2017A&A...605A..46M}, although its metallicity distribution is not as large as that in $\omega$~Cen and Terzan~5.

In recent years, several studies have been made with the objective to find other GCs displaying intrinsic iron spread; however, they present smaller inhomogeneities than $\omega$~Cen, Terzan~5, and M54.
Examples of such clusters are M22 \citep{2009A&A...505.1099M}, NGC~3201 \citep{2013ApJ...764L...7S}, M2 \citep{2014MNRAS.441.3396Y}, NGC~5286 \citep{2015MNRAS.450..815M}, and M19 \citep{2015AJ....150...63J}. Some of these works have been re-evaluated by other studies to either confirm the intrinsic spread in the metallicities or conclude that those findings were not correct at all.

NGC~3201 (C1015$-$461) is a nearby ($\sim 4.9$~kpc), low galactic latitude ($b = +8.64^\circ$) GC located at $\alpha = 10^\mathrm{h}~17^\mathrm{m}~36.^{\!\!{\mathrm{s}}}82$, 
$\delta = -46^{\circ}~24^{'}~44.^{\!\!{''}}9$ (J2000) \citep[2010 edition]{1996yCat.7195....0H}. Its $\text{[Fe/H]}$ content is controversial, since different analyzes have yielded notably different results. For example, the study of \citet{1998AJ....116..765G}, that uses  CTIO high-resolution spectra, and that of \citet{2013ApJ...764L...7S}, which employs high-resolution spectra from UVES-FLAMES@VLT and MIKE@Magellan, claim that NGC~3201 has a spread in $\text{[Fe/H]}$ of $\sim 0.4$~dex. However, \citet{2013MNRAS.433.2006M} do not find such spread, deriving $\text{[Fe/H]}=-1.53\pm0.01$~dex,  and  also point out that 5 out of the 6 most metal-poor stars---with $\text{[Fe/H]}\leq-1.59$~dex---from \citet{2013ApJ...764L...7S} might be asymptotic giant branch (AGB) stars instead of red giant branch (RGB) stars.

\citet{2015ApJ...801...69M} re-analyzed the spectra of 21 stars from the sample of \citet{2013ApJ...764L...7S}  in the light of the results of \citet{2014ApJ...797..124L}, who proposed non-local thermodynamical equilibrium (NLTE) effects as an explanation for the lower Fe abundances of AGB stars compared to those observed in RGB stars. By adopting photometric gravities that use the Stefan-Boltzmann equation, and measuring Fe~I and Fe~II lines individually, they discovered that the spread found in \citet{2013ApJ...764L...7S} was due to the presence of AGB stars in the sample, and concluded that NGC~3201 does not have an intrinsic iron spread. It should be noted, however, that for GCs with metallicity similar to NGC 3201, NLTE corrections models from \citet{2012MNRAS.427...27B} and \citet{2012MNRAS.427...50L} predict that the Fe~I lines should be affected in a similar way regardless the star is an RGB or AGB star, and thus there should be another mechanism that could explain the behavior of the iron content of the cluster.

In this context it is important to note that red giant stars are unstable against radial pulsations; in fact, as stars evolve expanding and cooling into the RG or AGB stages, they become pulsationally unstable, meaning that---virtually---every star found in the upper right portion of the H-R diagram is a variable. Moreover, pulsating red giants (PRGs) are complex objects, mainly due to their variability being the result of a mix of pulsation and convection (this not yet being a well understood process in astrophysics), and because of the presence of extended atmospheric envelopes \citep{2015pust.book.....C}. Olin Eggen published a series of papers during the 1970s (e.g., \citealt{1973ApJ...184..793E, 1977ApJ...213..767E}) in which he developed a general classification for PRGs: large, medium, and small-amplitude red variables (LARV, MARV, and SARV, respectively), along with the $\sigma$~Librae variables,  stars with amplitudes less than $0.2$~mag in $V$. Pulsating red giants are characterized by having a long period and a short amplitude.

As a first clue to a possible relation between variability and iron abundance in red giants, recently \citet{2018A&A...620A..96M} analyzed the chemical abundance of seven stars of NGC~6528, and  found that one of the RGB stars of the sample was: (1) more metal-poor, with $\mathrm{[Fe/H]}=-0.55$, than the other six stars---having  a mean value of $\mathrm{[Fe/H]}=-0.14\pm0.03$), and that (2) it was a variable star with a period of $0.26$~days and an infrared amplitude of $0.05$~mag, using data from the VVV survey \citep{2010NewA...15..433M, 2012A&A...537A.107S}. These findings prompted us to ask ourselves the following question: \textit{Can the variability of a star be related to its observed difference in iron abundance, compared with that of the rest of the stars that are not variable?} To seek for an answer, we decided to revisit one of the GCs whose iron content is debated, NGC~3201.
As a further motivation, it is worth pointing out  that \citet{2003AJ....125..208L} reported low-amplitude light variations in several RGB stars in this cluster.

In this paper we present a study of thirty-four light curves (LCs) of seventeen giant stars---17 in $V$ and 17 in $I$---of the globular cluster NGC~3201. First, we try to determine the possible variability of each star within the limitations of our data. Then for the possible variables, we then searched the  literature regarding their $\text{[Fe/H]}$ content. {The aim is to check if these giant stars are variables or not and if such variability is somehow related with their iron abundance as obtained from spectroscopic data using standard data analysis.} The layout of the paper is as follows. In Section~\S~\ref{sec2} we present the information regarding the data employed in the work. Section~\S~\ref{sec:3analisis} provides details of the methodology followed to analyze the sample. Section~\S~\ref{sec:resultados} contains the main results  obtained from our analysis. In Section~\S~\ref{sec:discuss} we present a discussion of our findings. Finally, in Section~\S~\ref{sec:conclus} we  summarize the results and give the conclusions obtained from this work.

\section{The Data}\label{sec2}
\begin{figure}
	\includegraphics[width=\textwidth]{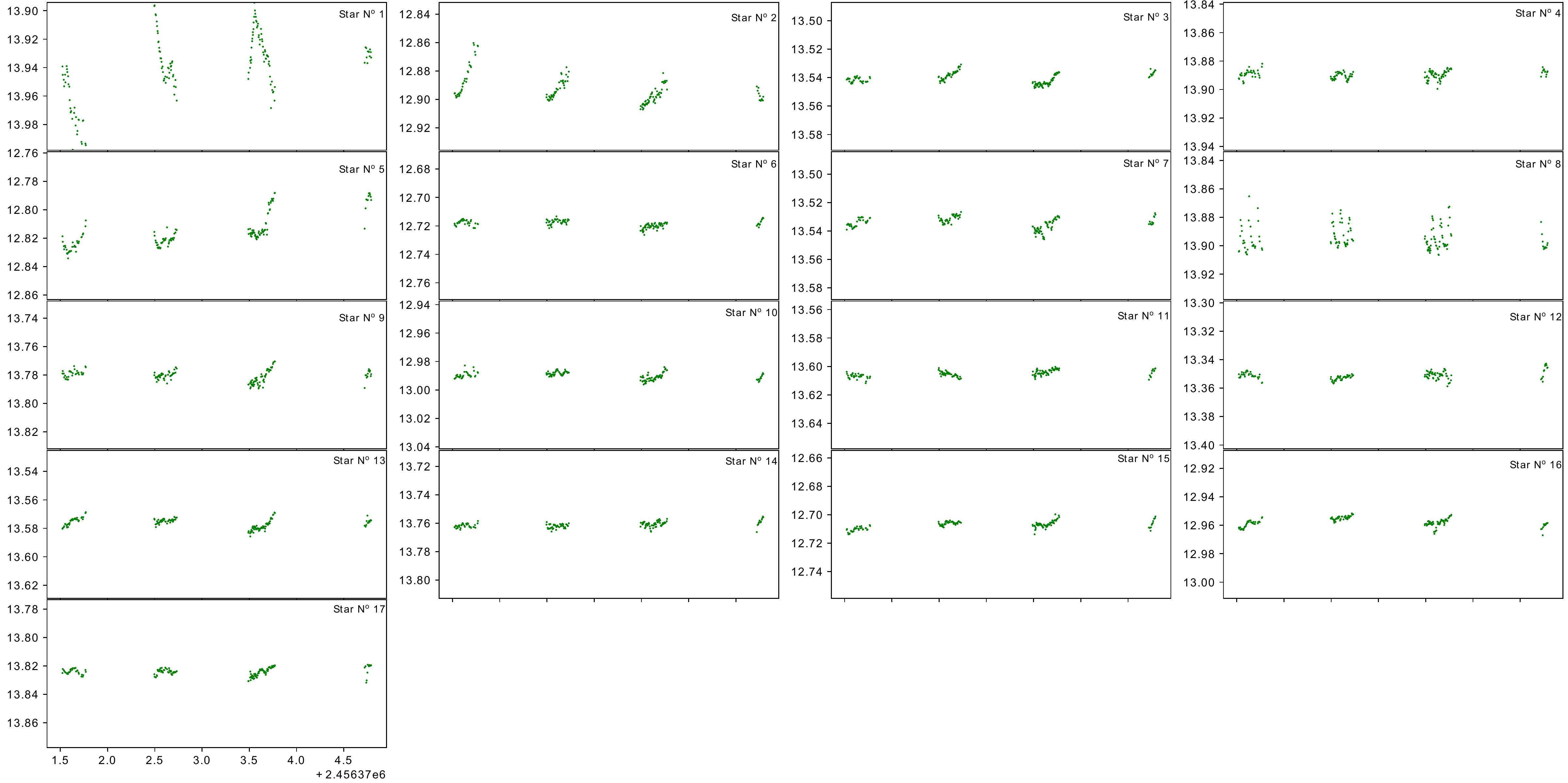}
    \caption{Light Curves of the 17 stars analyzed in this work for $V$ filter. The x-axis correspond to the Heliocentric Julian Days (HJD) and the y-axis shows the standard magnitude $V$. We named each RGB as `Star N$^{\mathrm{o}}$' followed by a number from 1 to 17 (e.g., Star N$^{\mathrm{o}}$ 1, Star N$^{\mathrm{o}}$ 2,..., Star N$^{\mathrm{o}}$ 17). The number is written in the top right corner of each LC, all light curves are plotted with the same scale.}
    \label{fig:LCsAllV}
\end{figure}

\begin{figure}
	\includegraphics[width=\textwidth]{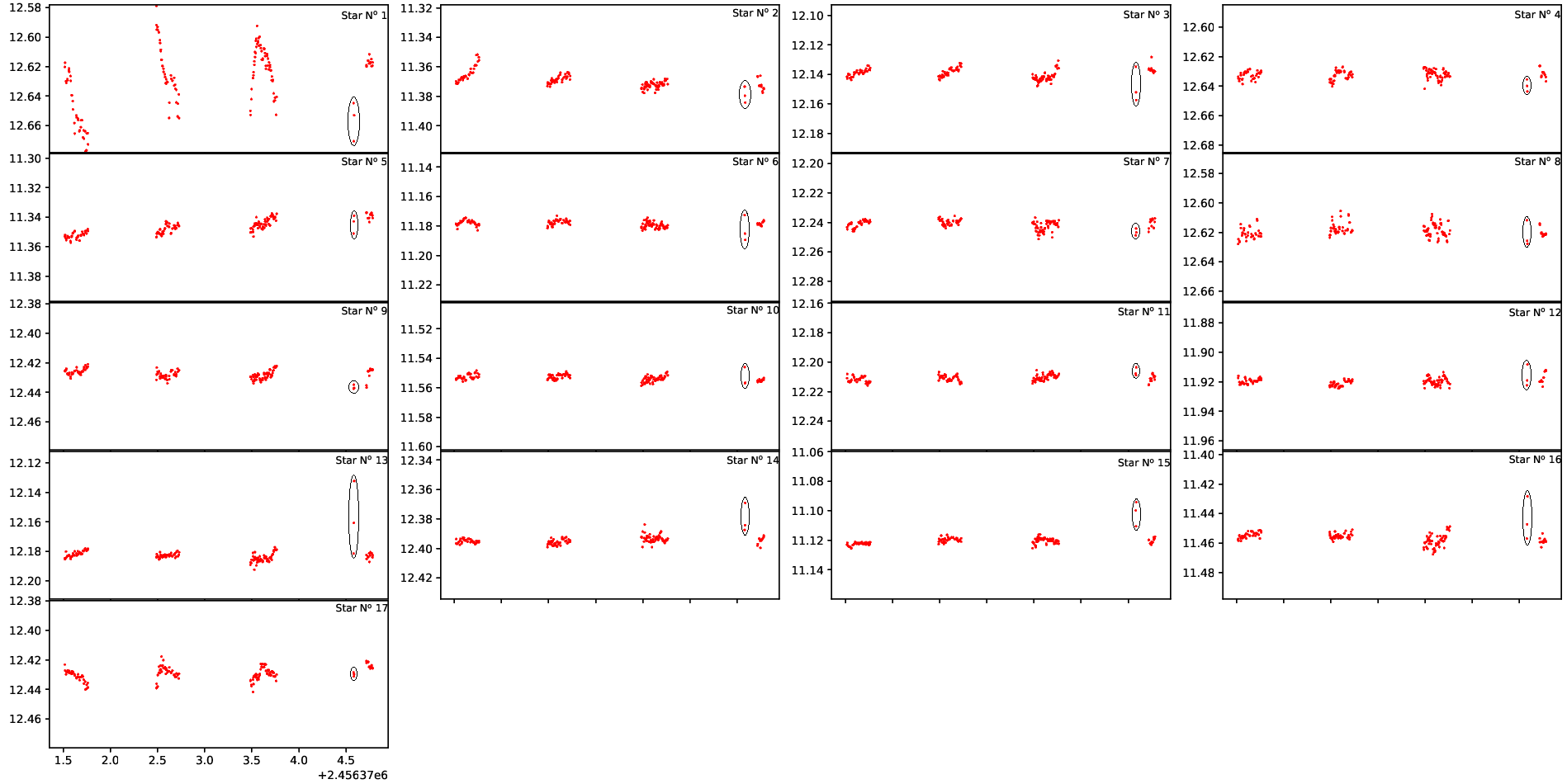}
    \caption{Light Curves of the 17 stars analyzed in this work for $I$ filter. Same as in Fig~\ref{fig:LCsAllV},  x-axis correspond to the Heliocentric Julian Days (HJD) and the y-axis shows the standard magnitude $I$. The black ellipses mark the 3 epoch that were affected by bad photometry and regarded as outliers.}
    \label{fig:LCsAllI}
\end{figure}

For this work we analyzed thirty-four light curves (17 for $V$ filter and 17 for $I$ filter) of seventeen giant stars from NGC 3201 provided by J.~Ahumada, previously published in \citet{2014RMxAA..50..307A}. These stars were also analyzed by \citet{2013ApJ...764L...7S} and \citet{2015ApJ...801...69M}, who determined  their $[\text{Fe/H}]$ abundance. Each light curve for the $V$ filter contains 142 epochs, and for the $I$ filter, four light curves (Stars N$^{\mathrm{o}}$ 2, 5, 6 and 15) are based on 144 epochs, while the other thirteen contain 145 epochs. Figure~\ref{fig:LCsAllV} and Figure~\ref{fig:LCsAllI} show the light curves of all stars for $V$ and $I$ filter, respectively, and Table~\ref{tab:Curvaejem} displays a portion of the complete content of each light curve.

\begin{table}[ht]
	\centering
	\caption{Sample time series for $V$ and $I$ filters of each star.}
	\label{tab:Curvaejem}
	\begin{tabular}{cccccc}
		\hline
		\textbf{Star}&\textbf{Filter}&\textbf{HJD}&\textbf{$M_{\mathrm{std}}$}&{\textbf{$\sigma_{M_{std}}$}}&\textbf{$\sigma_{int}$}\\
	&&\textbf{(d)}&\textbf{(mag)}&\textbf{(mag)}&\textbf{(mag)}\\
		\hline
	N$^{\mathrm{o}}$ 1&$V$&2456371.522859&13.93920&0.0078&0.00074\\
	N$^{\mathrm{o}}$ 1&$V$&2456371.525984&13.94512&0.0079&0.00063\\
	$\vdots$  &	$\vdots$  &	$\vdots$  & $\vdots$ & $\vdots$ & $\vdots$\\
	N$^{\mathrm{o}}$ 1&$I$&2456371.517222&12.62022&0.0092&0.00085\\
	N$^{\mathrm{o}}$ 1&$I$&2456371.519717&12.61734&0.0093&0.00080\\
	$\vdots$  &	$\vdots$  &	$\vdots$ & $\vdots$ & $\vdots$ & $\vdots$\\
	N$^{\mathrm{o}}$ 2&$V$&2456371.522859&12.89581&0.0090&0.00043\\
	N$^{\mathrm{o}}$ 2&$V$&2456371.525984&12.89554&0.0090&0.00035\\
	$\vdots$  &	$\vdots$  &	$\vdots$ & $\vdots$ & $\vdots$ & $\vdots$\\
	N$^{\mathrm{o}}$ 2&$I$&2456371.517222&11.36980&0.0095&0.00043\\
	N$^{\mathrm{o}}$ 2&$I$&2456371.519717&11.37122&0.0099&0.00041\\	
		\hline
	\end{tabular}
	\tablecomments{Col. 1 and 2 display the number of the star as in Figures~\ref{fig:LCsAllV} and~\ref{fig:LCsAllI} and the filter, respectively. Col. 3 shows the epoch of mid-exposure in heliocentric Julian days (HJD), col. 4 shows the standard magnitude, col. 5 are the errors associated to the standard magnitudes and col. 6 are the internal errors associated to the DanDIA reduction program.}
\end{table}

The data for all 17 giant stars were obtained on 2013 March 20--23 at the Complejo Astron\'{o}mico El~Leoncito (CASLEO), San Juan, Argentina, using the 2.15-m telescope, carrying out Johnson-Kron-Cousins $V$ and $I$ observations. The detector used was a Roper Scientific back-illuminated CCD of $2048\times2048$ pixels with a scale of $0.15$ arcsec/pix and a Field of View of approximately $5.1\times5.1$ arcmin$^2$. The data reduction and transformation to the \emph{VI} standard magnitude system are described  in Secs.~2.2 and 2.3 of \citet{2014RMxAA..50..307A}.

\begin{figure}
\centering
	\includegraphics[width=0.75\textwidth]{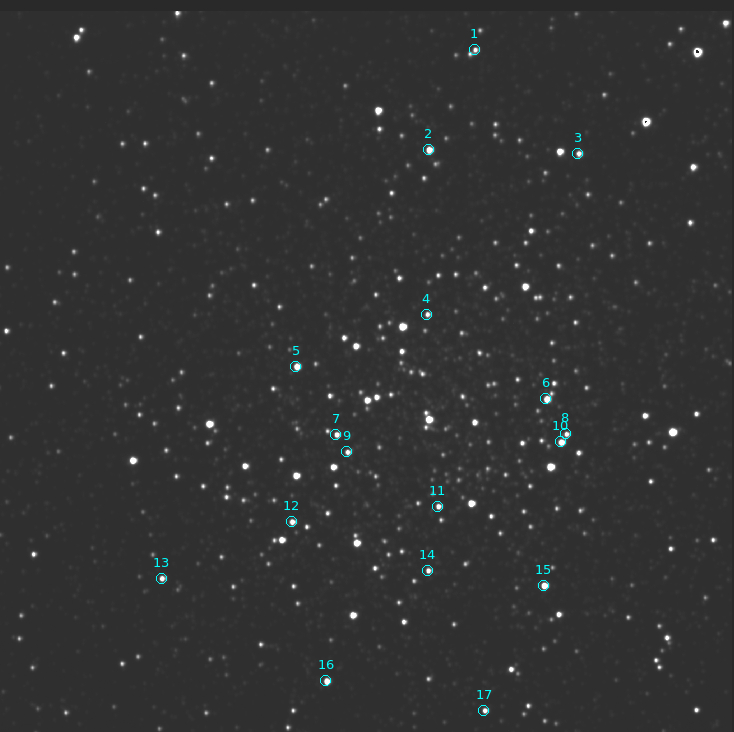}
    \caption{Finding chart of the area of NGC~3201 where the 17 giant stars analyzed in this work are indicated. These stars were studied previously by \citet{2013ApJ...764L...7S} and \citet{2015ApJ...801...69M}. West is up, North is to the left, and the shown area  is $5.4' \times 5.4'$.}
    \label{fig:Area}
\end{figure}

\begin{figure*}
    \includegraphics[width=\textwidth]{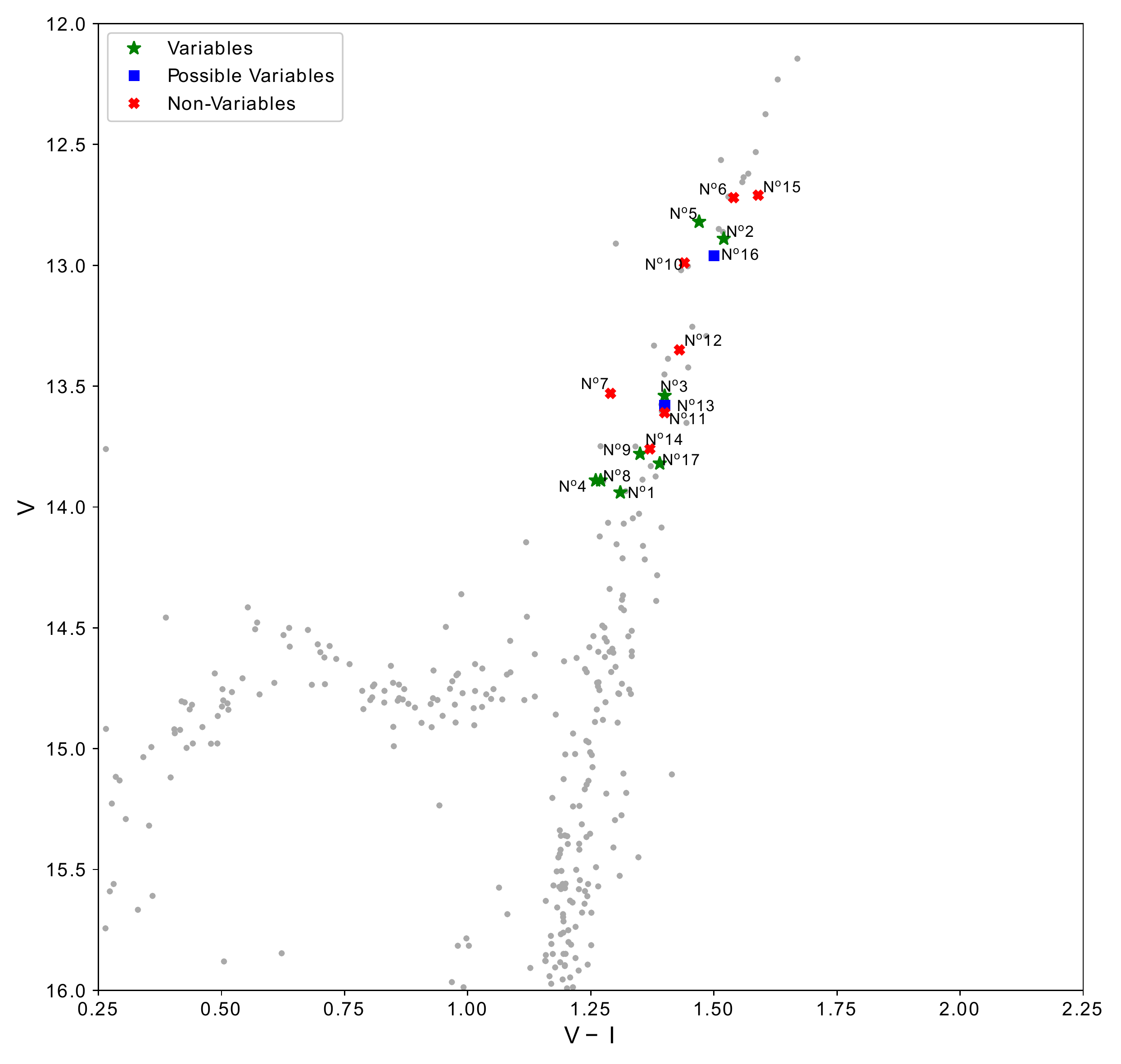}
    \caption{CMD of NGC 3201 displaying the 17 giant stars studied in this work with different markers based on their classification. Green stars are the variable stars, blue squares represent the possible variables, and the red crosses are the non-variables. Each star also has its corresponding ID.}
    \label{fig:CMDcluster}
\end{figure*}

The finding chart of the stars that were analyzed in this work is in Figure~\ref{fig:Area}. It is worth noting that Stars N$^{\mathrm{o}}$ 8 and 10 are close to each other, while Stars N$^{\mathrm{o}}$ 1, 6, 7, and 9 have close (although faint) neighbor stars. In spite of this, contamination from these neighbours on the spectra used to obtain [Fe/H] seems unlikely. This is because slit spectrographs like MIKE@Magellan can easily avoid neighbour contamination just rotating the camera and in any case during the reduction process the spectrum of a contaminating star can be successfully subtracted from that of the target. Fiber spectrograph like UVES-FLAMES@VLT have fibers that are only 1.2 arcsec in diameter and neighbours stars visible in Figure~\ref{fig:Area} are not close enough to affect the incoming light.
In order to identify our stars of interest, we matched them with the ones from \citet{2015ApJ...801...69M} using both position (RA and DEC) and mean magnitudes $\langle M_V\rangle$ and $\langle M_I\rangle$ for $V$ and $I$ filters, respectively. We also matched our stars with those from the variable star catalogue by \citet{2001AJ....122.2587C} using \textit{Tool for OPerations on Catalogues And Tables} (TOPCAT) to see if any of our stars were already catalogued as variable, where we did not find any match at all. This is not surprising since there are very few studies in literature about variability in globular cluster giant branch stars. Table~\ref{tab:datosstars} shows the main information of the each stars and Figure~\ref{fig:CMDcluster} shows the Color-Magnitude Diagram (CMD) of the cluster with the position of the stars studied in this work marked with different symbols depending if they were classified as variable, non-variable, or possible variable, as explained in \S~\ref{sec:resultados}.

\begin{table*}[ht]
	\centering
	\caption{Basic information of the 17 giant stars studied in this work.}
	\label{tab:datosstars}
	\begin{tabular}{lcccccccc} 
		\hline
		Star & RA & DEC & $\langle M_V\rangle$ & $\langle M_I\rangle$ & $[\text{Fe/H}]$& $[\text{Fe I/H}]$& $[\text{Fe II/H}]$\\
		N$^{\mathrm{o}}$ & (J2000) & (J2000) & (mag) & (mag) & (dex) & (dex)& (dex)\\
		\hline
		1 & $154.3433125$ & $-46.4197750$ & $13.94$ & $12.63$ &$-1.48$& $-1.31\pm0.03$&$-1.29\pm0.05$\\
		2 & $154.3596208$ & $-46.4137333$ & $12.89$ & $11.37$ &$-1.42$& $-1.37\pm0.02$&$-1.40\pm0.05$\\
		3 & $154.3617667$ & $-46.4310361$ & $13.54$ & $12.14$ &$-1.54$& $-1.48\pm0.02$&$-1.43\pm0.04$\\
		4 & $154.3872500$ & $-46.4123778$ & $13.89$ & $12.63$ &$-1.79$& $-1.62\pm0.03$&$-1.37\pm0.04$\\
		5 & $154.3946708$ & $-46.3968944$ & $12.82$ & $11.35$ &$-1.53$&$-1.50\pm0.02$&$-1.40\pm0.04$\\
		6 & $154.4027208$ & $-46.4256111$ & $12.72$ & $11.18$ &$-1.45$& $-1.47\pm0.02$&$-1.45\pm0.05$\\
		7 & $154.4065292$ & $-46.4010306$ & $13.53$ & $12.24$ &$-1.65$& $-1.54\pm0.02$&$-1.39\pm0.03$\\
		8 & $154.4087875$ & $-46.4276361$ & $13.89$ & $12.62$ &$-1.71$& $-1.61\pm0.03$&$-1.38\pm0.04$\\
		9 & $154.4095667$ & $-46.4021667$ & $13.78$ & $12.43$ &$-1.56$& $-1.44\pm0.02$&$-1.44\pm0.05$\\
		10 & $154.4101042$ & $-46.4269639$ & $12.99$ & $11.55$ &$-1.64$& $-1.56\pm0.02$&$-1.42\pm0.04$\\
		11 & $154.4196083$ & $-46.4122944$ & $13.61$ & $12.21$ &$-1.50$& $-1.40\pm0.02$&$-1.42\pm0.05$\\
		12 & $154.4207125$ & $-46.3952389$ & $13.35$ & $11.92$ & $-1.54$& $-1.43\pm0.03$&$-1.43\pm0.04$\\
		13 & $154.4288708$ & $-46.3798167$ & $13.58$ & $12.18$ &$-1.51$& $-1.45\pm0.02$&$-1.45\pm0.04$\\
		14 & $154.4302625$ & $-46.4106583$ & $13.76$ & $12.39$ & $-1.52$& $-1.50\pm0.02$&$-1.45\pm0.04$\\
		15 & $154.4340000$ & $-46.4239750$ & $12.71$ & $11.12$ &$-1.46$ & $-1.44\pm0.03$&$-1.44\pm0.06$\\
		16 & $154.4478125$ & $-46.3981083$ & $12.96$ & $11.46$ &$-1.59$& $-1.54\pm0.02$&$-1.38\pm0.04$\\
		17 & $154.4543542$ & $-46.4162194$ & $13.82$ & $12.43$ &$-1.45$& $-1.31\pm0.03$&$-1.29\pm0.05$\\
		\hline
	\end{tabular}
	\tablecomments{First, second and third columns give the ID of the stars and their coordinates. Fourth and fifth the mean $V$ and $I$ magnitudes, sixth column gives the $[\text{Fe/H}]$ values are from \citet{2013ApJ...764L...7S}, while seventh and eighth columns give the $[\text{Fe I/H}]$ and $[\text{Fe II/H}]$ values are from \citet{2015ApJ...801...69M}.}
\end{table*}

%%%%%%%%%%%%%%%%%%%%%%%%%%%%%%%%%%%%%%%%%%%%%%%%%%
%%%%%%%%%%%%%%%%%%%%%%%%%%%%%%%%%%%%%%%%%%%%%%%%%%
\section{Analysis of the Light Curves and Search for Variability}\label{sec:3analisis}
To check the possible variability of each star, we applied both the \textit{Generalized Lomb-Scargle} (GLS) \citep{2009A&A...496..577Z} and the \textit{Phase Dispersion Minimization}  \citep{1978ApJ...224..953S} methods. We must first note that, for all stars observed  in the $I$ filter, there are 3 epochs taken in the fourth night, which are indicated with an ellipse in Figure~\ref{fig:LCsAllI}. These epochs show evidence of having been affected by bad photometry, and were therefore rejected for our analysis on the grounds that they are outliers.

Both analyzes were performed in \textit{Python} using the \textit{PyAstronomy}\footnote{\url{https://github.com/sczesla/PyAstronomy}} (PyA) package collection \citep{2019ascl.soft06010C}. The code requires as input:  observation time, the data (in this case magnitudes), and (optionally) the errors associated with them (cols. 3, 4, and 5 of Table~\ref{tab:Curvaejem}, respectively), along with a Starting Period ($P_{\mathrm{beg}}$) and an End Period ($P_{\mathrm{end}}$).

The frequency step utilized for the analysis is given by the equation:
\begin{equation}
    {f_{\mathrm{step}} = 1.0/\mathrm{(Time~Length)}/\mathrm{OFac},}
\end{equation}
with Time Length being HJD$_{\mathrm{end}}-\mathrm{HJD}_{\mathrm{beg}}$, and OFac is the Oversampling Factor, which has a value of 10. The GLS analysis also carries out a False-Alarm-Probability (FAP), used to check which periods are significant enough. The FAP is given by $$\mathrm{FAP}(P_n)= 1-{(1-\mathrm{Prob}(P>P_n))}^M,$$ where $M$ is the number of independent power values, computed internally by the program and $\mathrm{Prob}()$ is the probability that a periodogram power $P$ has to exceed a given value $P_n$. This function depends on the periodogram and on the normalization.

\begin{figure*}
  \includegraphics[width=\textwidth]{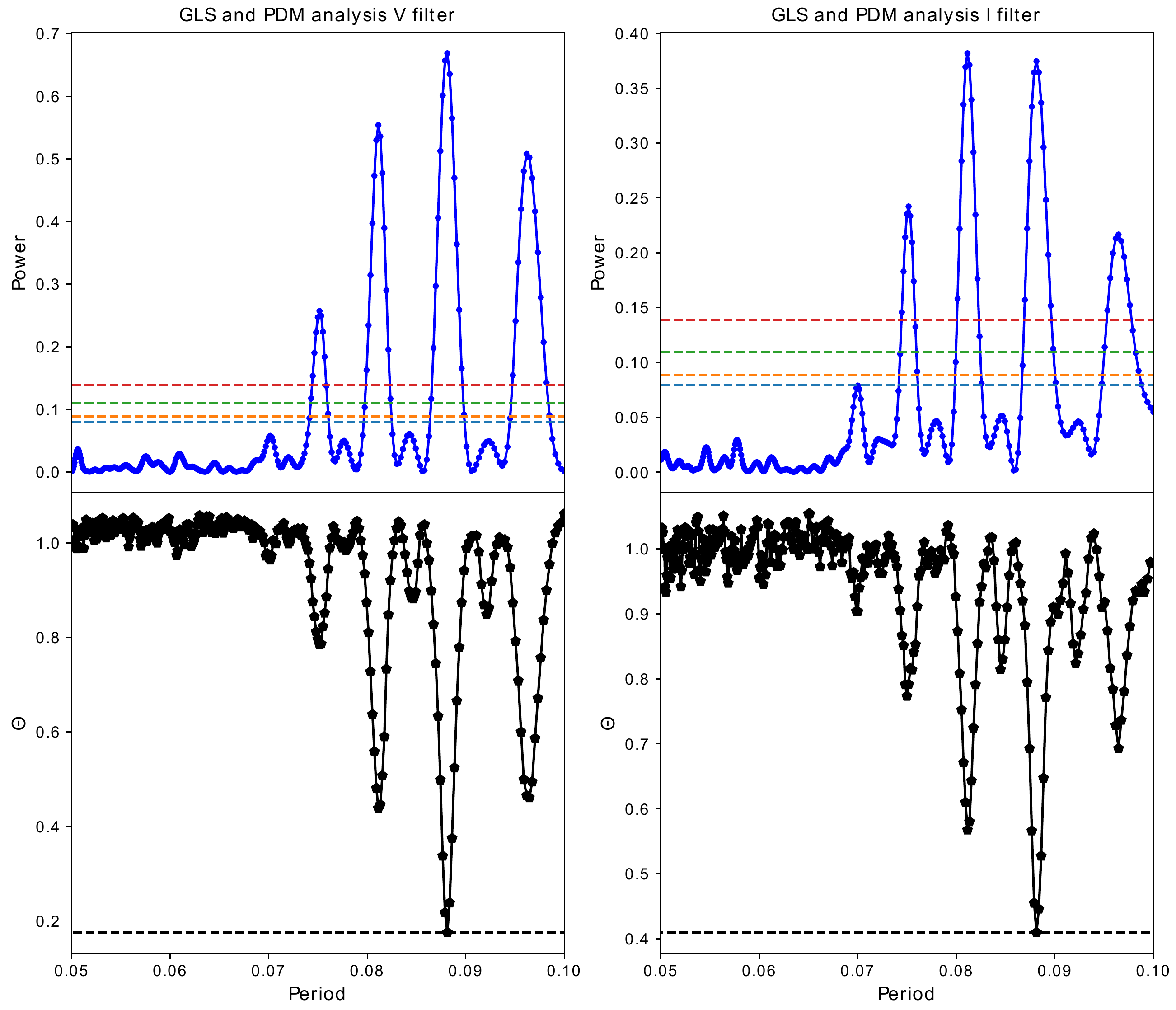}
    \caption{Resulting plots from the GLS (top panels) and PDM (bottom panels) analysis performed with the \textit{PyAstronomy} package collection for $V$ and $I$ filters (left and right panels, respectively). Both results are from the initial period analysis of Star N$^{\mathrm{o}}$8. From the GLS analysis one can see the different FAP level thresholds indicated with the different colored dashed lines, with blue being 10\%, yellow 5\%, green 1\% and red 0.1\% probability. In the PDM analysis the black dashed line indicates the minimum value of $\Theta$. Note that the GLS analysis for $I$ filter is the only one that marks a different period ($P=0.0811$~d) compared to the ones marked by GLS in $V$ filter and PDM for both filters.}
    \label{fig:GLSPDMtest}
\end{figure*}

After performing the GLS analysis, the program carried out the PDM analysis for the star using the same parameters as before, viz., Starting Period, End Period, and the Frequency Step. Figure~\ref{fig:GLSPDMtest} shows an example of the  periodograms obtained for Star N$^{\mathrm{o}}$~8 with the GLS (top panels) and PDM (bottom panels) methods, with the left and right panels being the results for $V$ and $I$ filters, respectively.

In order to check if a giant of our sample can be considered as a candidate variable star or not, we adopted the following criteria:
\begin{enumerate}
    \item The inequality $A\pm \sigma_A > \text{rms}_{\mathrm{mag}}$ must be satisfied for both filters, with $A$, $\sigma_{A}$, and $\mathrm{rms}_{\mathrm{mag}}$ being the amplitude, its error, and the root-mean square of the magnitude data, respectively. These parameters are given by the GLS analysis.
    
    \item The possible periods must display their peaks above the False Alarm Probability (FAP) threshold of 0.1\% in the GLS periodogram for both filters.
    
    \item The periods $P_{\text{GLS}}$ and $P_{\text{PDM}}$ must be similar for both filters.
    
    \item The possible period must be present in both  filters, and $P_V\approx P_I$.
    
    \item The visual inspection of the light curves must show that they maintain a constant mean magnitude along the four nights of observation.
\end{enumerate}
Adopting these criteria, the stars discarded {as non-variables} are N$^{\mathrm{o}}$s 6, 7, 10, 12, 14, and 15.

%%%%%%%%%%%%%%%%%%%%%%%%%%%%%%%%%%%%%%%%%%%%%%%%%%
\subsection{Significance Test}\label{subsec:significancias}
The purpose of this section is to present how we determined which giant could be a candidate variable or not, as well as to determine the degree of variability of each star through a \textit{significance test}.

For this purpose we first calculate the magnitude difference $\Delta M$ for each star, defined as
\begin{equation}
    \Delta M = M - \langle M\rangle,
    \label{eq:delta}
\end{equation}
with $\langle M\rangle$ being the mean magnitude of the star. As in the preliminary analysis, the three bad data points of the fourth night in the $I$ filter (indicated with ellipses) affected the measurement of $\langle M_I\rangle$ and, in consequence, $\Delta M_I$, and therefore were discarded.

After obtaining  $\Delta M$ for all 17 giant stars, we first calculated $\sigma_M = \sqrt{\frac{\sum{\Delta M^2}}{N-1}}$ and its error $\mathrm{err}_\sigma = \frac{\sigma_ M}{{\sqrt{2N}}}$ for the six objects that were discarded as non-variable, using the data of all those stars combined {together} for each filter, which we named $\sigma_{\text{Non Var.}}$ and $\mathrm{err}_{\sigma_{\text{Non Var.}}}$. {This represents the variance of a typical non-variable star}. Then, we proceeded to calculate $\sigma_{\mathrm{Star}}$ and {$\mathrm{err}_{\sigma_{\mathrm{Star}}}$} for the remaining eleven candidate variables, to determine their significance $S$ (for each filter), given by 
\begin{equation}
    S=\frac{\sigma_{\mathrm{Star}}-\sigma_{\text{Non Var.}}}{\sqrt{\mathrm{err}_{\sigma_{\mathrm{Star}}}^2 + \mathrm{err}_{\sigma_{\text{Non Var.}}}^2}}.
    \label{eq:significancia}
\end{equation}
To find out whether or not the star could be a variable, we adopted the following criteria based on the value of  $S$ (Eq.~\ref{eq:significancia}):
\begin{enumerate}
    \item If $S<2$ for both filters, the star is not considered a variable.
    
    \item If $S>3$ for both filters, the star is considered a variable.
    
    \item If $S>3$ in one of the filters, the star is considered a possible variable.
    
    \item If $2< S < 3$ in one or both filters, the star is considered a dubious variable and requires further analysis for confirmation.
\end{enumerate}
For stars  in the cases (ii), (iii), and (iv), we re-performed the period analysis described in the previous section by carefully adjusting the different parameters from the PyAstronomy routines, while stars in the case (i) were not re-analysed. Table~\ref{tab:signif} shows the values of $S$ derived for all candidate variables along with the classification according to the above criteria. One star---N$^{\mathrm{o}}$s 11---was rejected as  variables due to its low values of $S_V$ and $S_I$, which are below 2. Six stars---N$^{\mathrm{o}}$s 1, 2, 3, 5, 8, and 9---were regarded as variables, since $S_V>3$ and $S_I>3$ for all of them, and their LCs from Figure~\ref{fig:LCsAllV} and Figure~\ref{fig:LCsAllI} show a clear change in their magnitudes each night. Four stars were considered as possible variables (N$^{\mathrm{o}}$s 4, 13, 16, and  17), and were re-analyzed to see if they could be variables, or candidates for future investigations. No star was classified as dubious variable.

\begin{table}
	\centering
	\caption{Significance $S$ values for $V$ and $I$ filters for the eleven preliminary candidate variable stars and their classification based on the criteria from Section~\ref{subsec:significancias}.}
	\label{tab:signif}
	\begin{tabular}{lccc} 
		\hline
		\textbf{Star} & $S_V$ & $S_I$ & Classification\\
		\hline
N$^{\mathrm{o}}$ 1&{14.719}&{14.874}&Variable\\
N$^{\mathrm{o}}$ 2&{11.695}&{8.311} &Variable\\
N$^{\mathrm{o}}$ 3&{3.524}& {4.885} & {Variable}\\
N$^{\mathrm{o}}$ 4&{0.490}&{3.900}& {Possible}\\
N$^{\mathrm{o}}$ 5&{12.648}&{8.596}&Variable\\
N$^{\mathrm{o}}$ 8&{11.013}&{7.833}&Variable\\
N$^{\mathrm{o}}$ 9&{4.800}&{3.162}&{Variable}\\
N$^{\mathrm{o}}$ 11&{0.000}&{0.000}&Non-Variable\\
N$^{\mathrm{o}}$ 13&{2.454}&{0.990}&{Possible}\\
N$^{\mathrm{o}}$ 16&{0.880}&{5.769}&Possible\\
N$^{\mathrm{o}}$ 17&{0.000}&{7.570}&Possible\\
		\hline
	\end{tabular}
\end{table}

%%%%%%%%%%%%%%%%%%%%%%%%%%%%%%%%%%%%%%%%%%%%%%%%%%
%%%%%%%%%%%%%%%%%%%%%%%%%%%%%%%%%%%%%%%%%%%%%%%%%%
\section{Analysis of the sample of giants}\label{sec:resultados}
In this section we {repeated the period analysis of the ten stars that were considered variable or possible variable} according to  the results listed in Table~\ref{tab:signif} (\S~\ref{subsec:significancias}). The final classification of our targets is reported in Table~\ref{tab:TablaAbundancias}. For the analysis of the variable stars, we have considered their characteristics, position in the CMD (Figure~\ref{fig:CMDcluster}) and comparing their light curves with existing variable stars.

\subsection{Variable Stars}\label{subsec:variables}
\begin{figure*}
\gridline{\fig{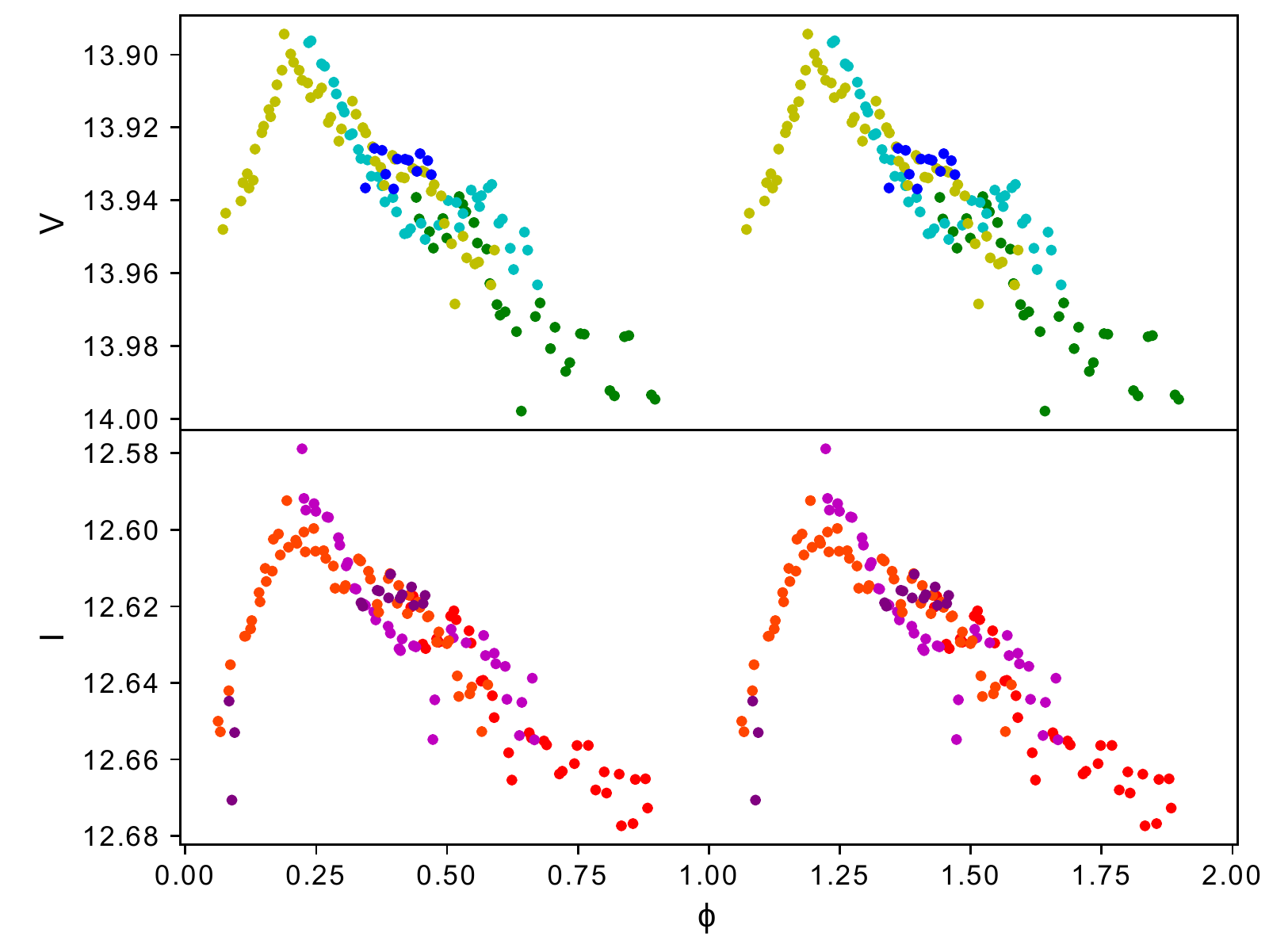}{0.33\textwidth}{(a) Star N$^{\mathrm{o}}$ 1}
          \fig{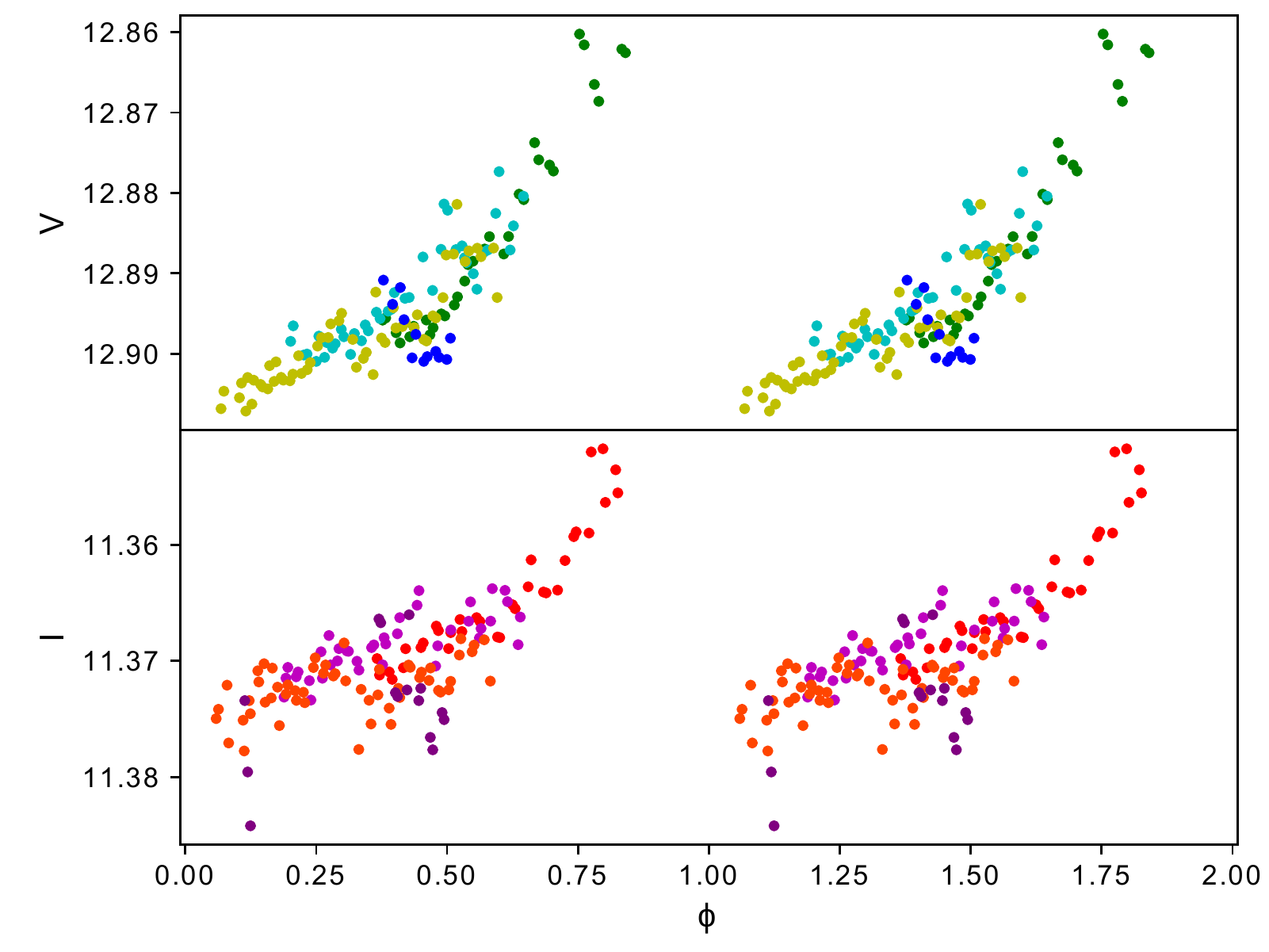}{0.33\textwidth}{(b) Star N$^{\mathrm{o}}$ 2}
          \fig{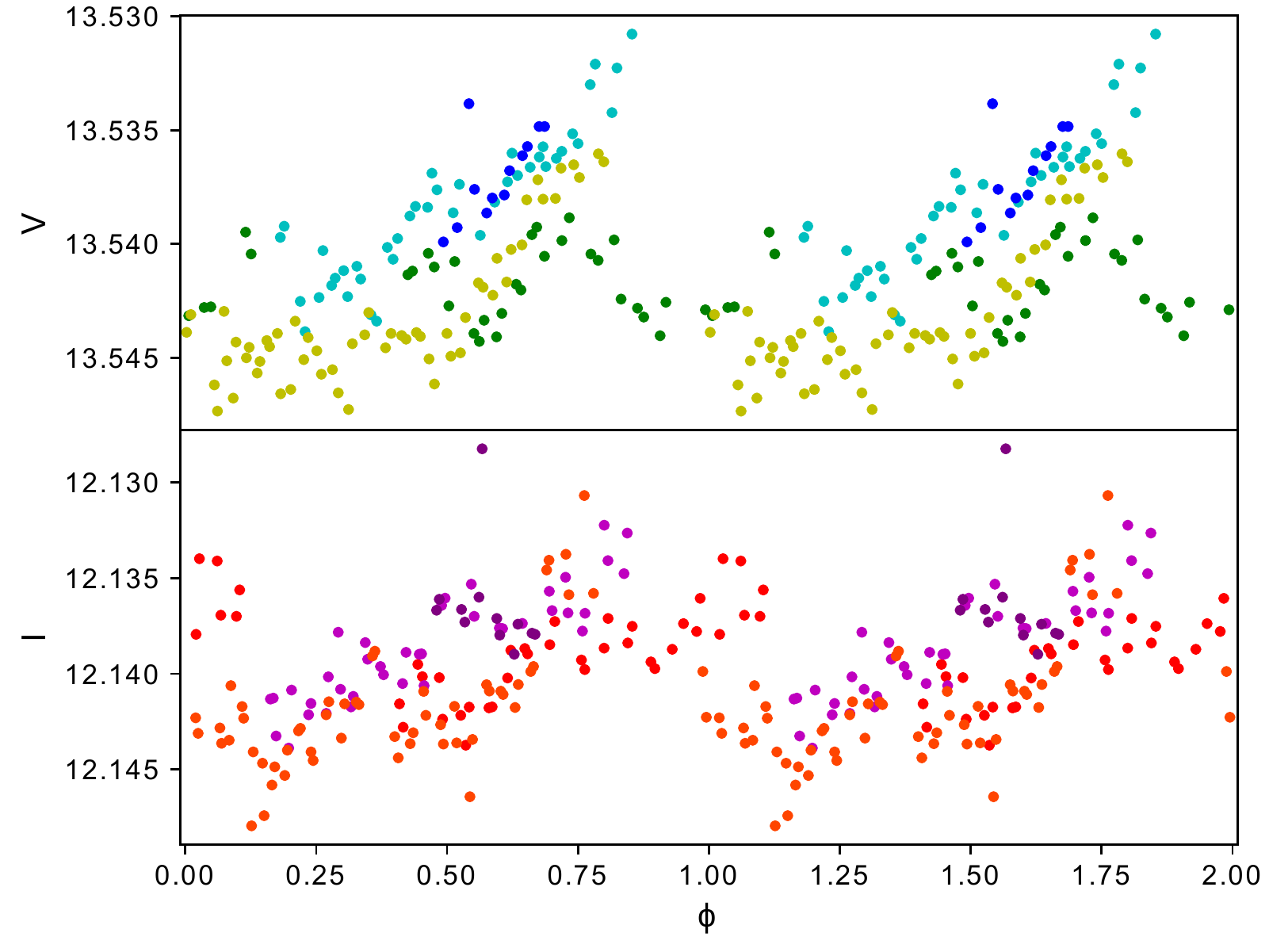}{0.33\textwidth}{(c) Star N$^{\mathrm{o}}$ 3}\label{fig:phaseS2}
          }
\gridline{\fig{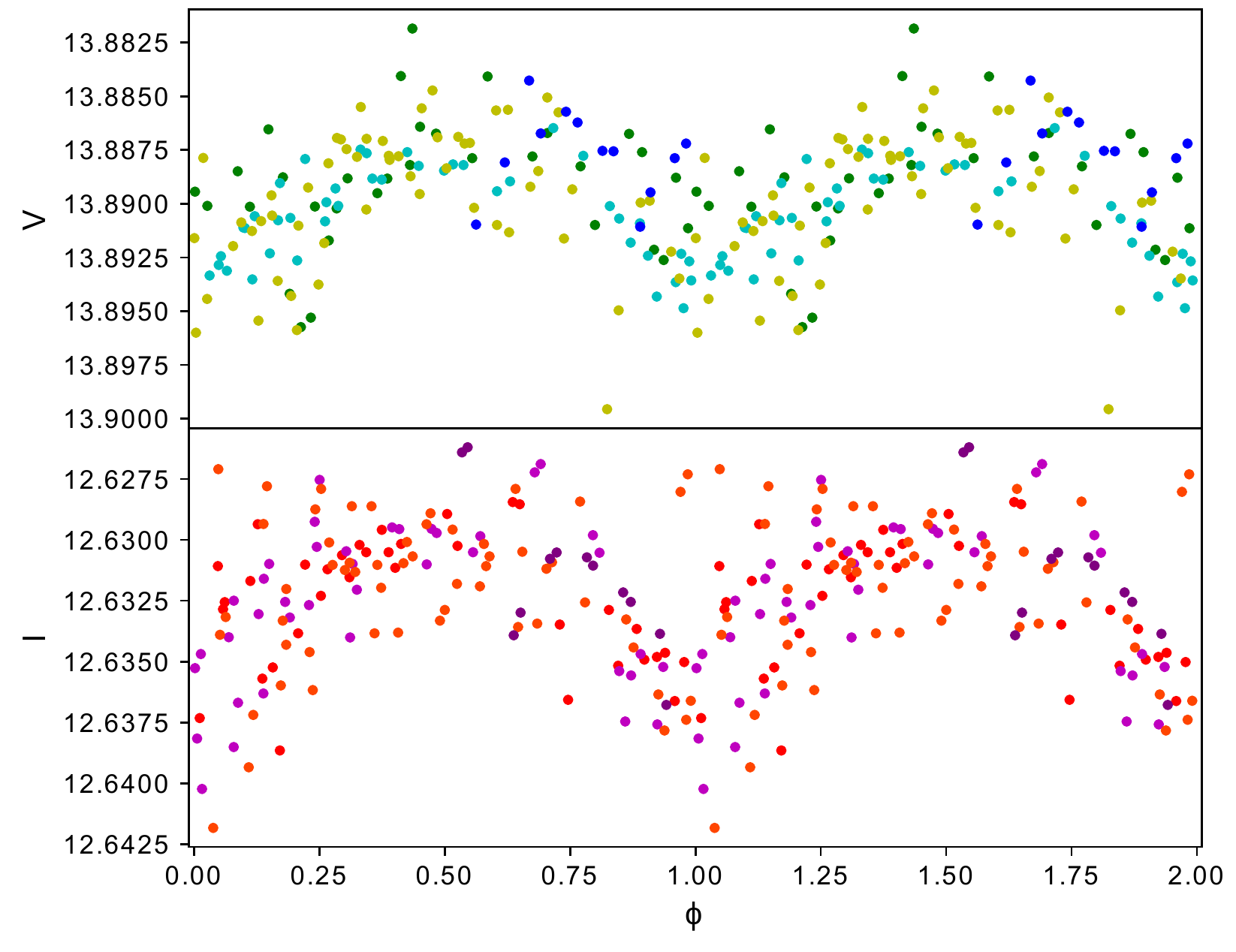}{0.33\textwidth}{(d) Star N$^{\mathrm{o}}$ 4}\label{fig:phaseS4}
\fig{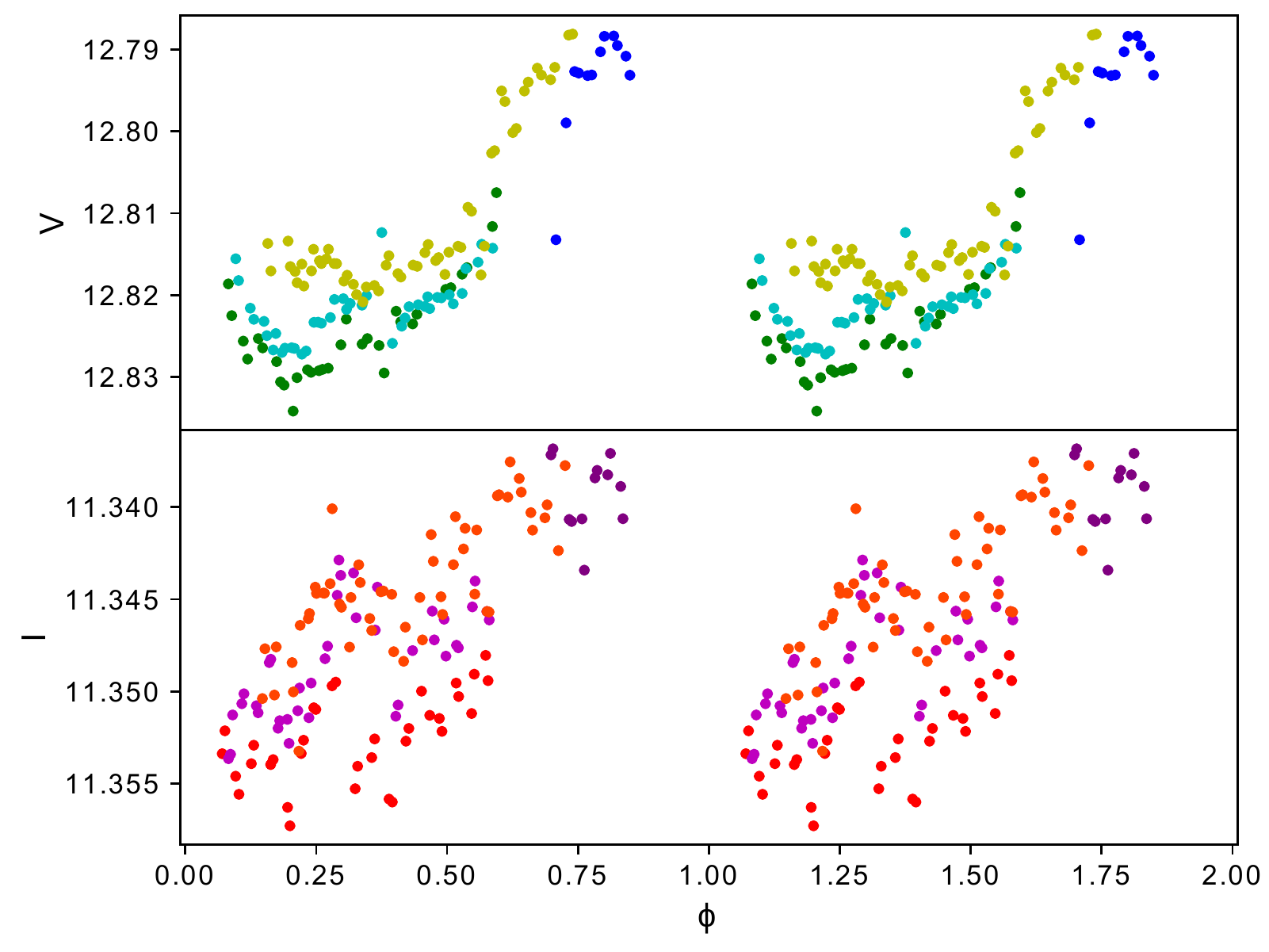}{0.33\textwidth}{(e) Star N$^{\mathrm{o}}$ 5}\label{fig:phaseS5}
          \fig{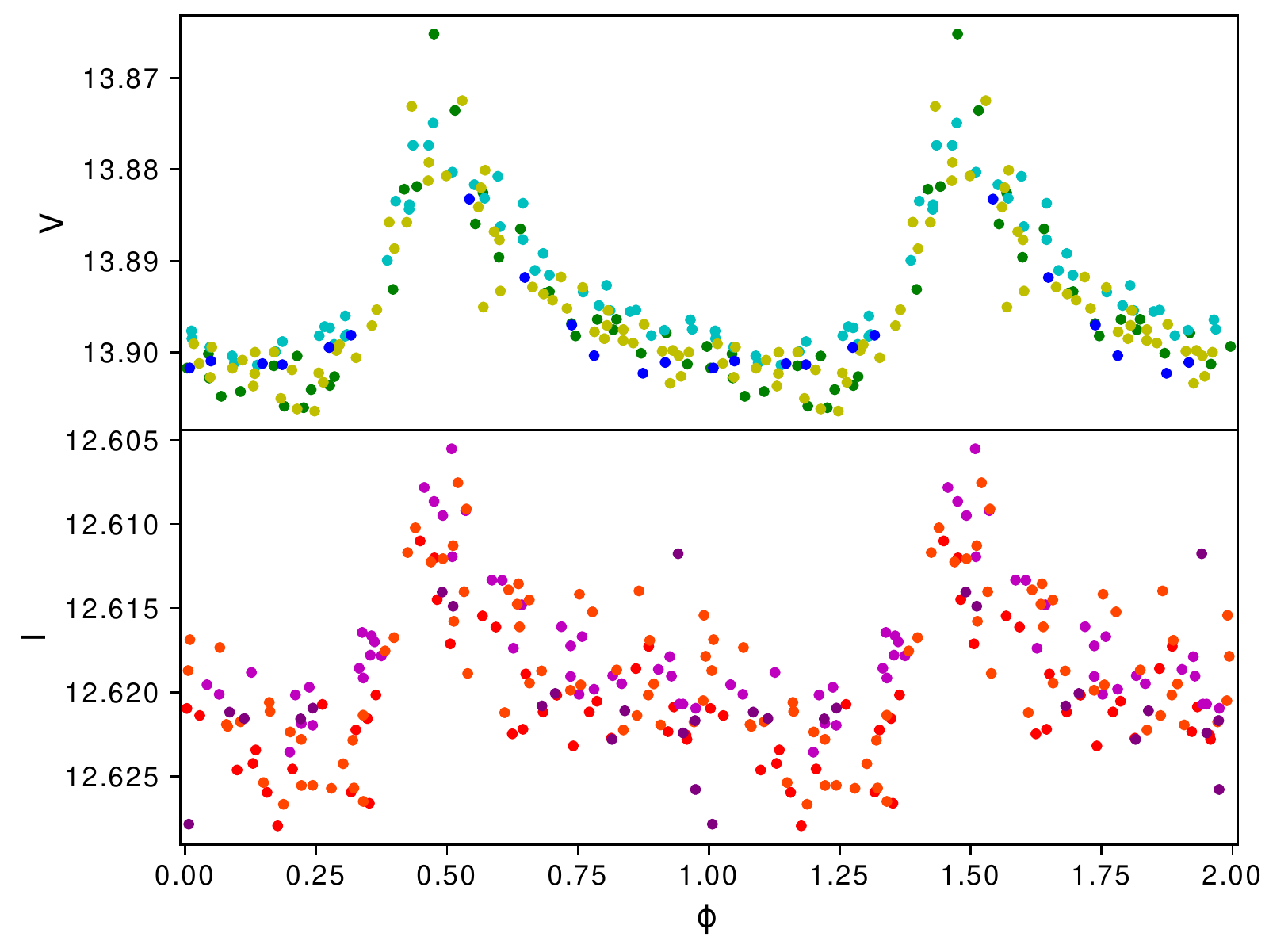}{0.33\textwidth}{(f) Star N$^{\mathrm{o}}$ 8}\label{fig:phaseS17}
          }
\gridline{\fig{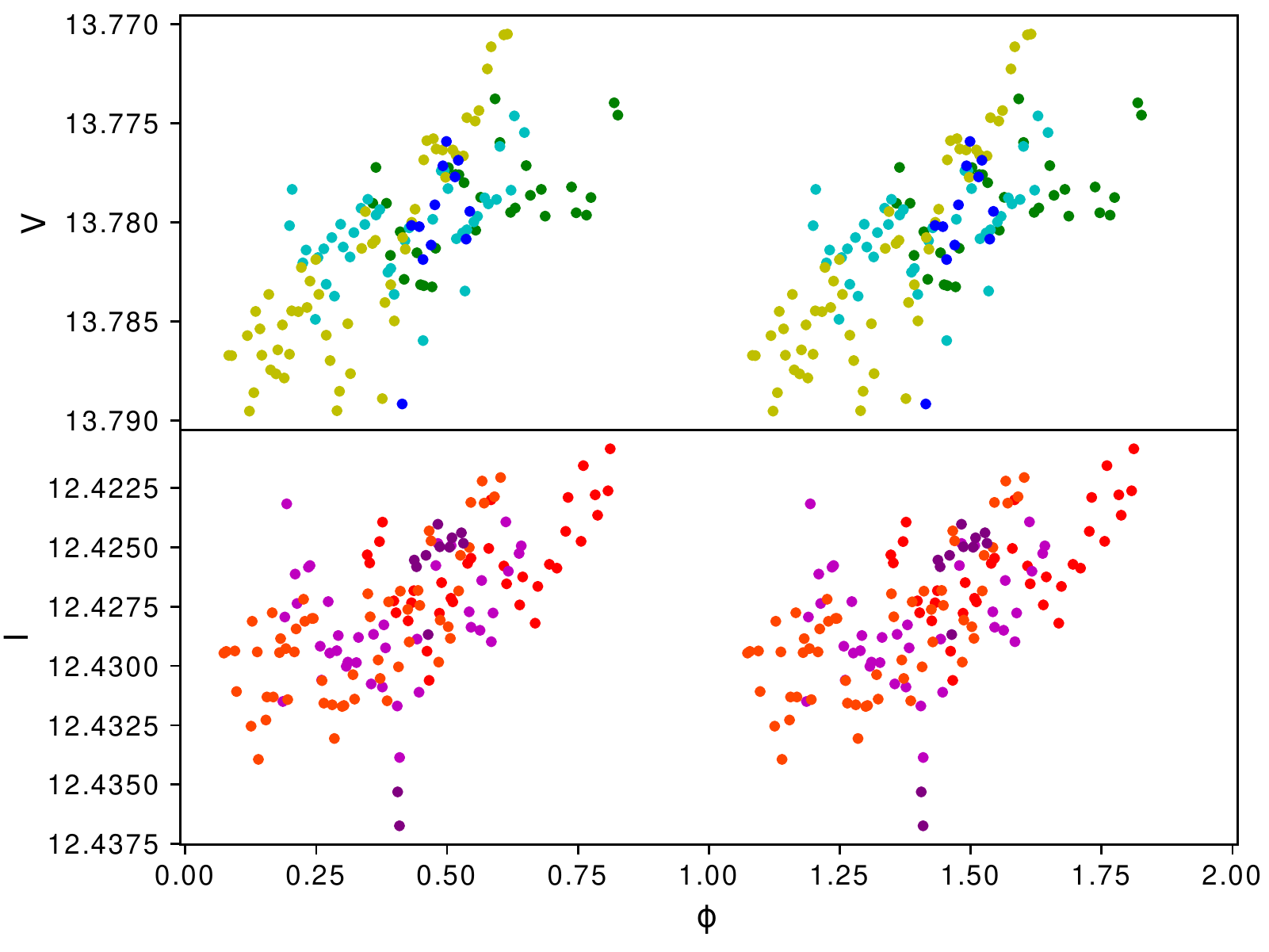}{0.33\textwidth}{(g) Star N$^{\mathrm{o}}$ 9}
          \fig{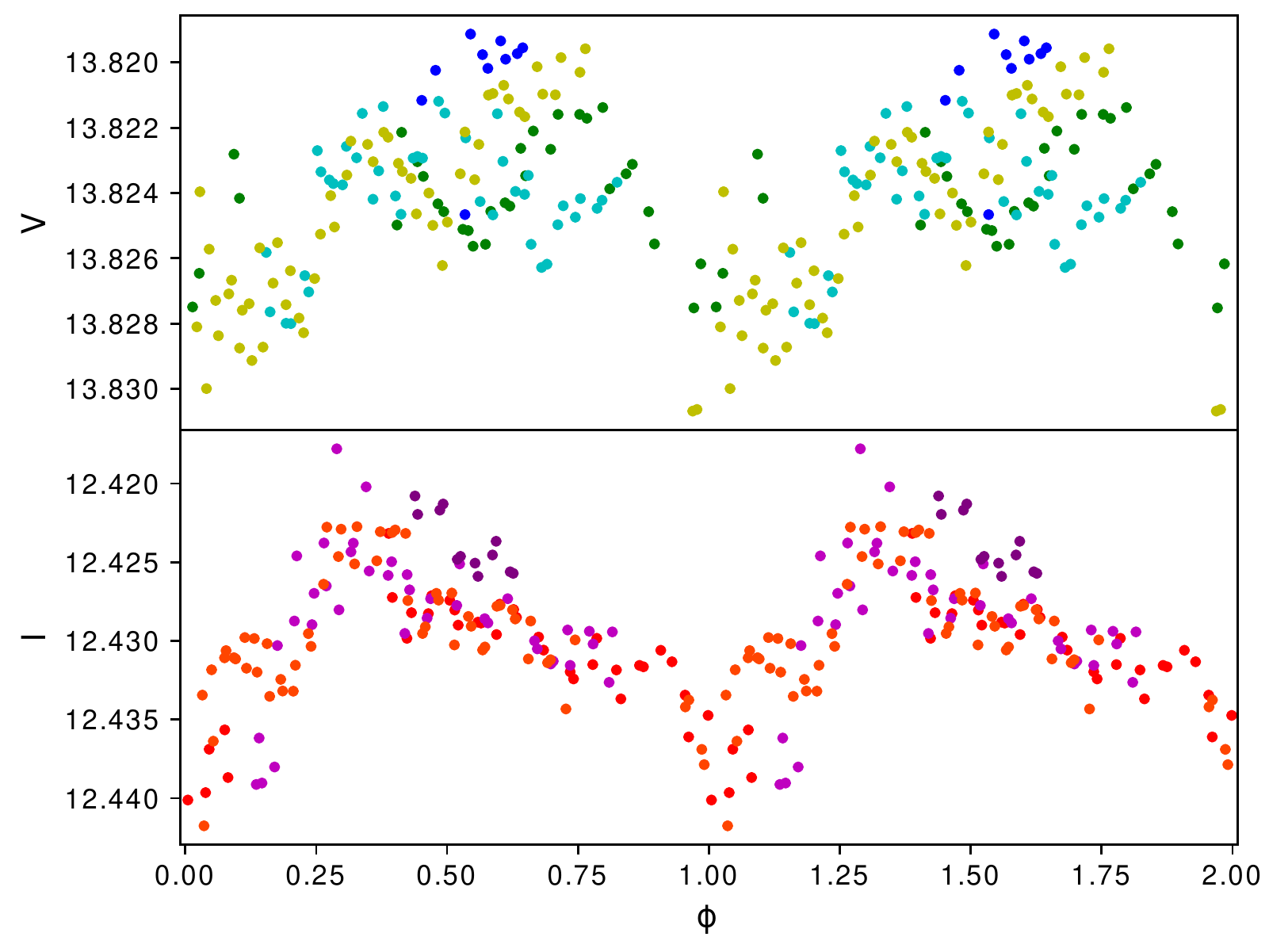}{0.33\textwidth}{(h) Star N$^{\mathrm{o}}$ 17}
          }

\caption{Phased light curves of our 8 variable stars found in this work which were plotted by phasing the LC of each star with their corresponding period $P_V$ shown in Table~\ref{tab:Results}. The top panel of each phased light curve correspond to the phased curve in the $V$ filter and the bottom panel the phased curve in the $I$ filter. The different colors indicate the data taken during each night of observation. For the top panel Night 1 is green, Night 2 cyan, Night 3 yellow and Night 4 blue. For the bottom panel Night is 1 red, Night 2 magenta, Night 3 orange and Night 4 purple.}
\label{fig:plotstodos}
\end{figure*}

In this section we will describe in detail each variable star found in this work. We applied again to the stars N$^{\mathrm{o}}$ 1, 2, 3, 4, 5, 8, 9, 13, 16, and 17 the period analysis described in \S~\ref{sec:3analisis} and we could confirm that stars 4 and 17 are also variable. Table~\ref{tab:Results} details the parameters for all the confirmed variable stars, and Figures~\ref{fig:plotstodos} shows their phased light curves for both filter and each night is represented by a different color. We will also compare their LCs with known variables taken as reference from the Optical Gravitational Lensing Experiment (OGLE) Atlas of Variable Star Light Curves\footnote{http://ogle.astrouw.edu.pl/atlas/index.html} \citep{1997AcA....47..319U, 2005AcA....55...43S, 2008AcA....58...69U} in order to attempt a classification. We consider also \citet{2018RMxAA..54...15Y} where the authors study variability in the GC NGC~6934 and identify a kind of variable in the giant branch of the cluster named semi regular (SR), which LC resemble some of the LCs of our targets. We underline the fact that in some case the reference best match variables have different parameters (mass, temperature, and gravity) compared with our targets, and also are located at different evolutionary phase. In any case this comparison can give us important hints about the physical mechanism behind the variability of our targets. We notice first that confirmed variable two kind of LCs. The first, represented by stars 1, 8 and 17, shows a sudden rise in magnitude followed by a slow drop. The second, represented by stars 2, 3, 4, 5 and 9, shows the opposite behavior, a slow rise followed by a fast drop.

\begin{table*}[ht]
	\centering
	\caption{Periodicity parameters for all stars that presented a variable behavior.}
	\label{tab:Results}
	\begin{tabular}{lcccccccccc} 
		\hline
		\textbf{Star} & $P_V$ & $\sigma_{P_V}$ & $A(V)$  & $\sigma_{A(V)}$ & $P_{I}$ & $\sigma_{P_I}$ & $A(I)$ & $\sigma_{A(I)}$ & $\langle M_V\rangle$ & $\langle M_I\rangle$\\
		 & (days)& (days) & (mag) & (mag) & (days) & (days) & (mag) & (mag) & (mag) & (mag)\\
		\hline
		N$^{\mathrm{o}}$ 1& 0.5418 & 0.0027 & 0.0591 & 0.0025 & 0.5417 & 0.0023 & 0.0556 & 0.0021 & 13.9393& 12.6293\\
		N$^{\mathrm{o}}$ 2& 0.5329 & 0.0026 & 0.0266 & 0.0012 & 0.5417 & 0.0043 & 0.0116 & 0.0008 & 12.8925& 11.3689\\
		{N$^{\mathrm{o}}$ 3}&{0.3527}&{0.0037}&{0.0064}&{0.0007}&{0.3429}&{0.0034}&{0.0050}&{0.0006}&{13.5405}&{12.1394}\\
		N$^{\mathrm{o}}$ 4& 0.1628 & 0.0008 & 0.0047 & 0.0005 & 0.1628 & 0.0007 & 0.0053 & 0.0006 & 13.8897& 12.6326\\
		N$^{\mathrm{o}}$ 5& 0.4827 & 0.0024 & 0.0331 & 0.0013 & 0.4772 & 0.0034 & 0.0114 & 0.0009 &12.8158& 11.3473\\
		N$^{\mathrm{o}}$ 8& 0.0881 & 0.0001 & 0.0195 & 0.0012 & 0.0881 & 0.0002 & 0.0072 & 0.0004 & 13.8944 & 12.6193\\
		{N$^{\mathrm{o}}$ 9}&{0.5281}&{0.0054}&{0.0081}&{0.0006}&{0.5164}&{0.0067}& {0.0058}&{0.0005}&{13.7802}&{12.4272}\\		
		N$^{\mathrm{o}}$ 17& 0.3535 & 0.0035 & 0.0045 & 0.0005 & 0.3535 & 0.0021 & 0.0093 & 0.0007 & 13.8238 & 12.4290\\
		\hline
	\end{tabular}
\end{table*}

\textbf{Star N$^{\mathrm{o}}$ 1:} This star has a period  $P=0.5418\pm0.0027$ days, with an amplitude $A>0.055$~mag in both filters. It shows a clear variable behavior in $V$ and $I$ (Figures~\ref{fig:LCsAllV} and~\ref{fig:LCsAllI}) and with a clear phased light curve (Figure~\ref{fig:plotstodos}\textcolor{blue}{a}). At first, for $V$ filter the GLS analysis showed us a period of $P\sim1.1135$ days if we considered a $P_{\mathrm{end}} = \text{Time Length}$, but the PDM analysis marked $P\sim0.55996$~d, while in the $I$ filter both the GLS and PDM analysis displayed a period between $0.5413$ and $0.55457$~d. The longest period obtained from the GLS analysis of the $V$ filter is almost twice the period obtained from the PDM analysis of the same filter and from the GLS and PDM analysis of the $I$ filter, and it could have been a considered as the actual period. However the phased light curve of this star displayed with such period was not realistic, which lead us to discard it. Finally, after restricting the $P_{\mathrm{end}}$ value to $P<1$~day, we obtained the period listed in Table~\ref{tab:Results}. It is worth noting that this star is positioned in the lower part of the RGB (see Figure~\ref{fig:CMDcluster}), and comparing its phased LC with known variables, it resembles an RR Lyrae ab star and the period is similar to some of the known RRab. However, its low amplitude does not agree with these type of variables.

\begin{figure*}
\gridline{\fig{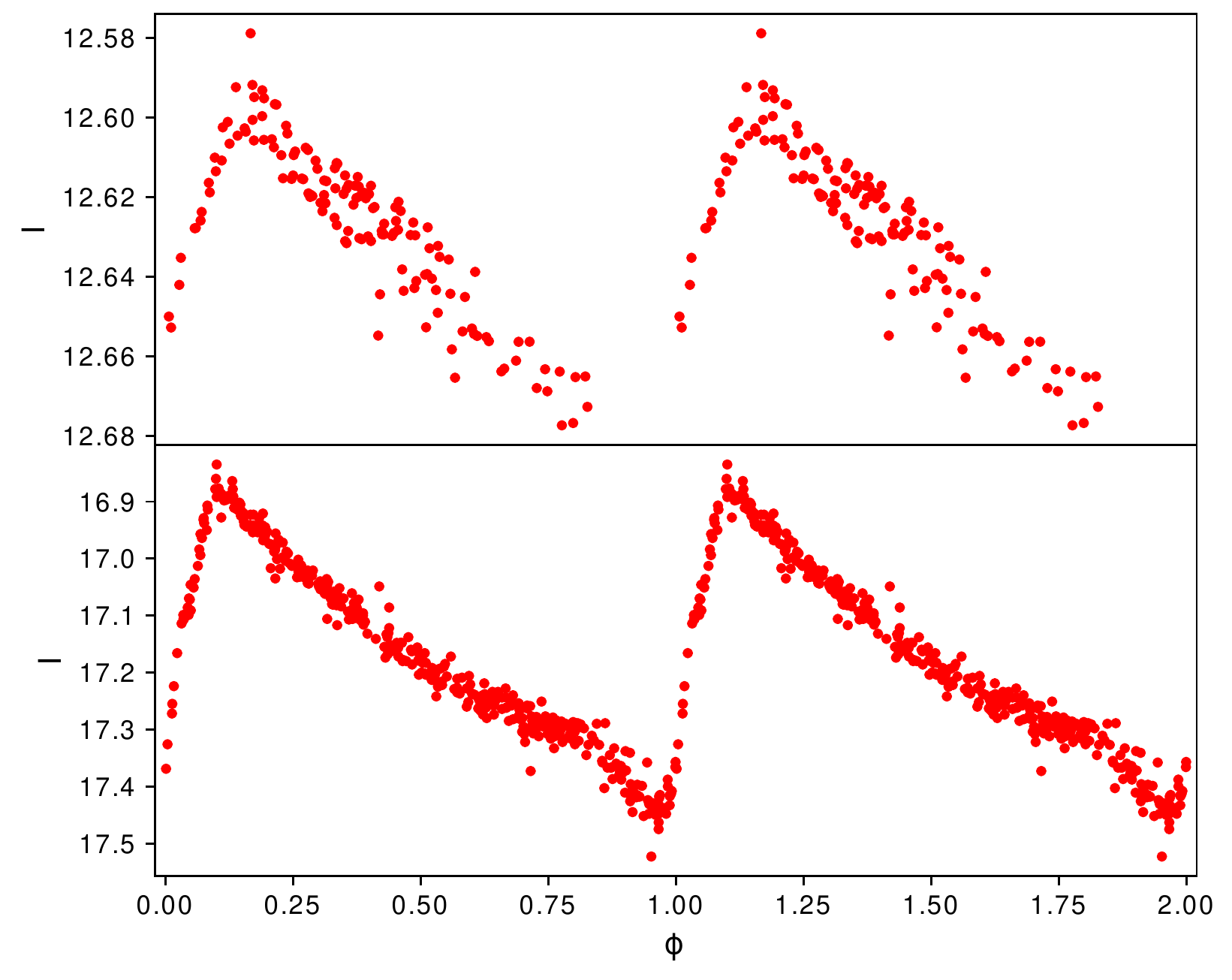}{0.33\textwidth}{(a) Star N$^{\mathrm{o}}$ 1 vs RR Lyrae ab OGLE-BLG-RRLYR-01004}
          \fig{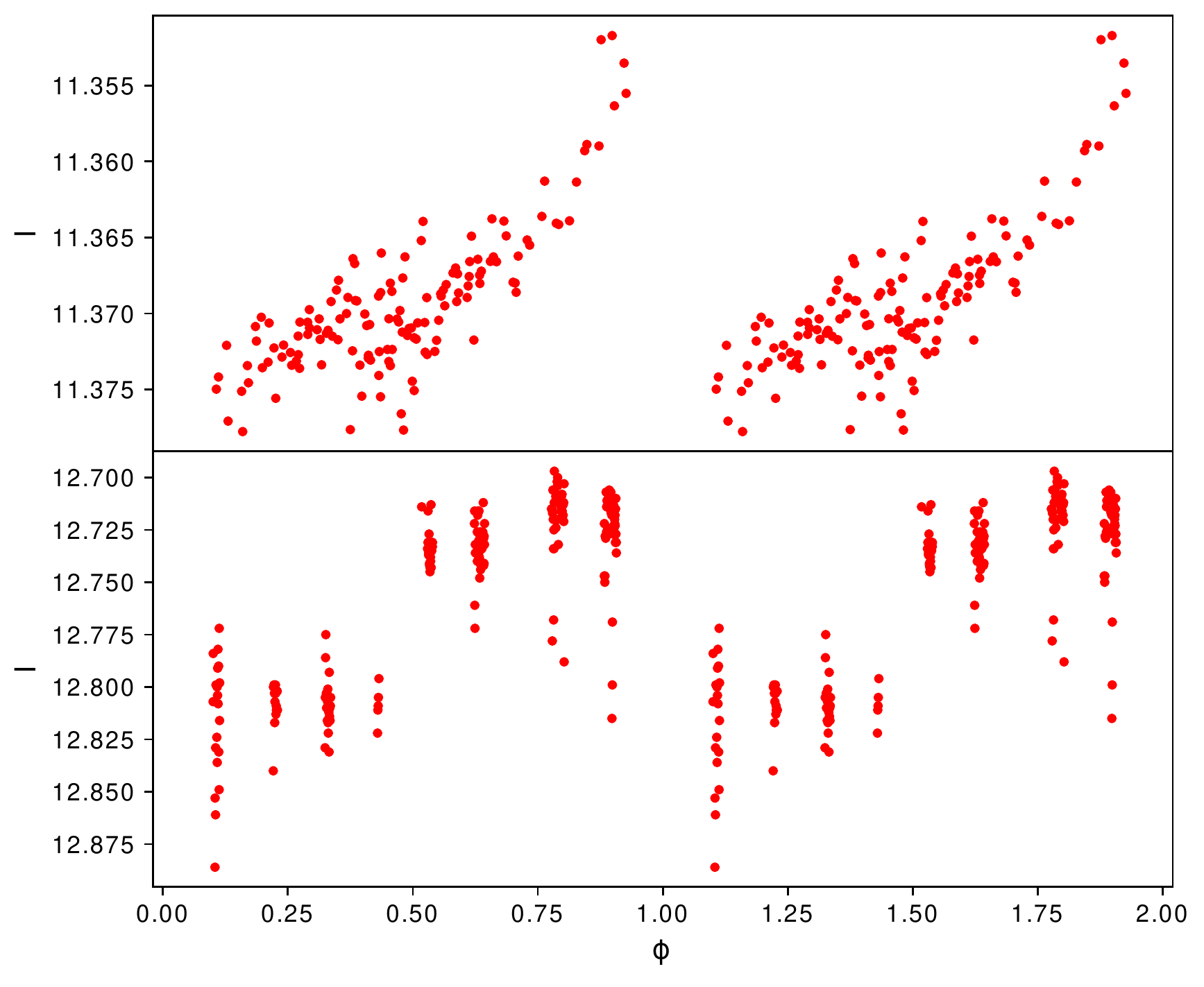}{0.33\textwidth}{(b) Star N$^{\mathrm{o}}$ 2 vs V96}
          \fig{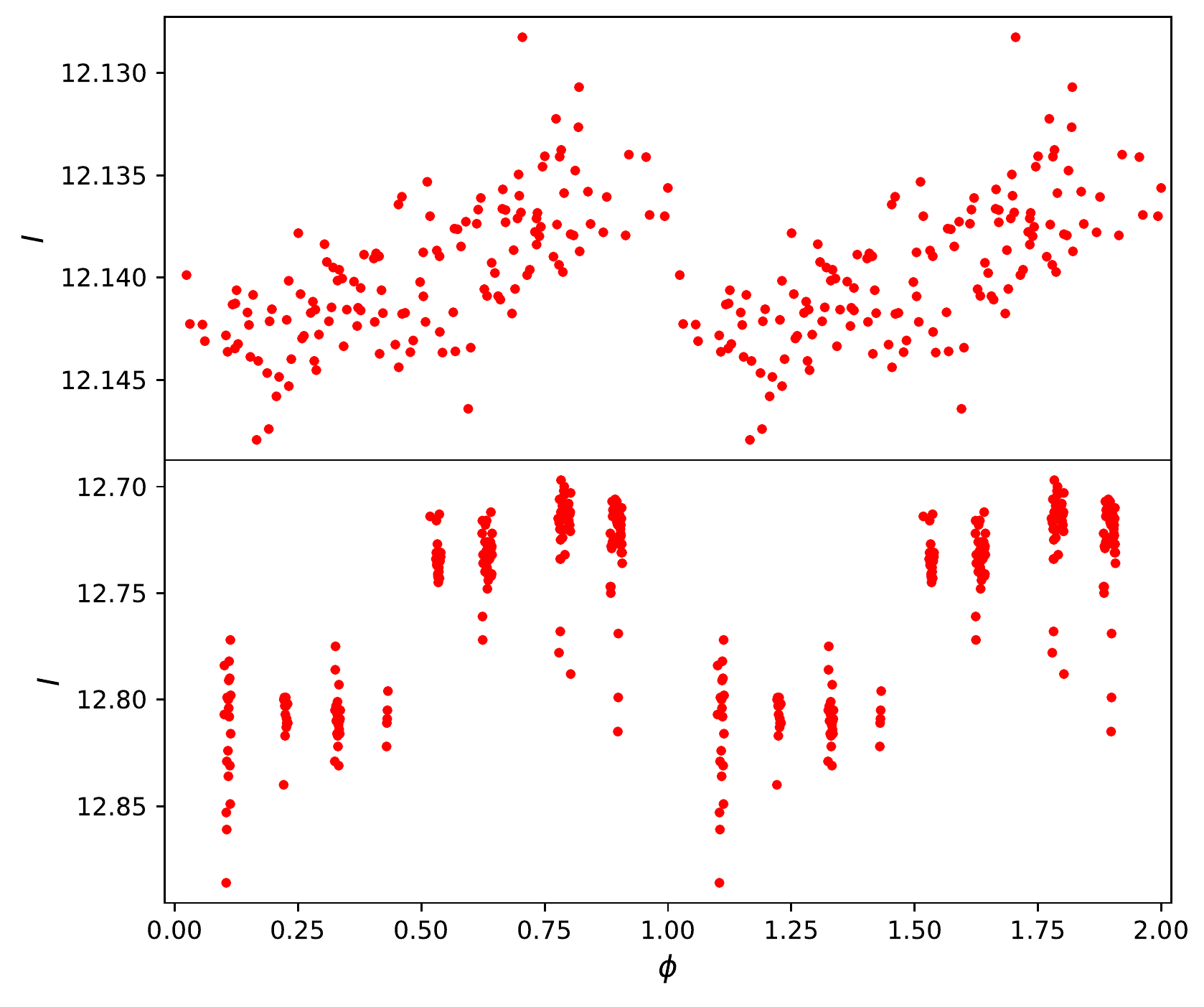}{0.33\textwidth}{(c) Star N$^{\mathrm{o}}$ 3 vs V96}
          }
\gridline{\fig{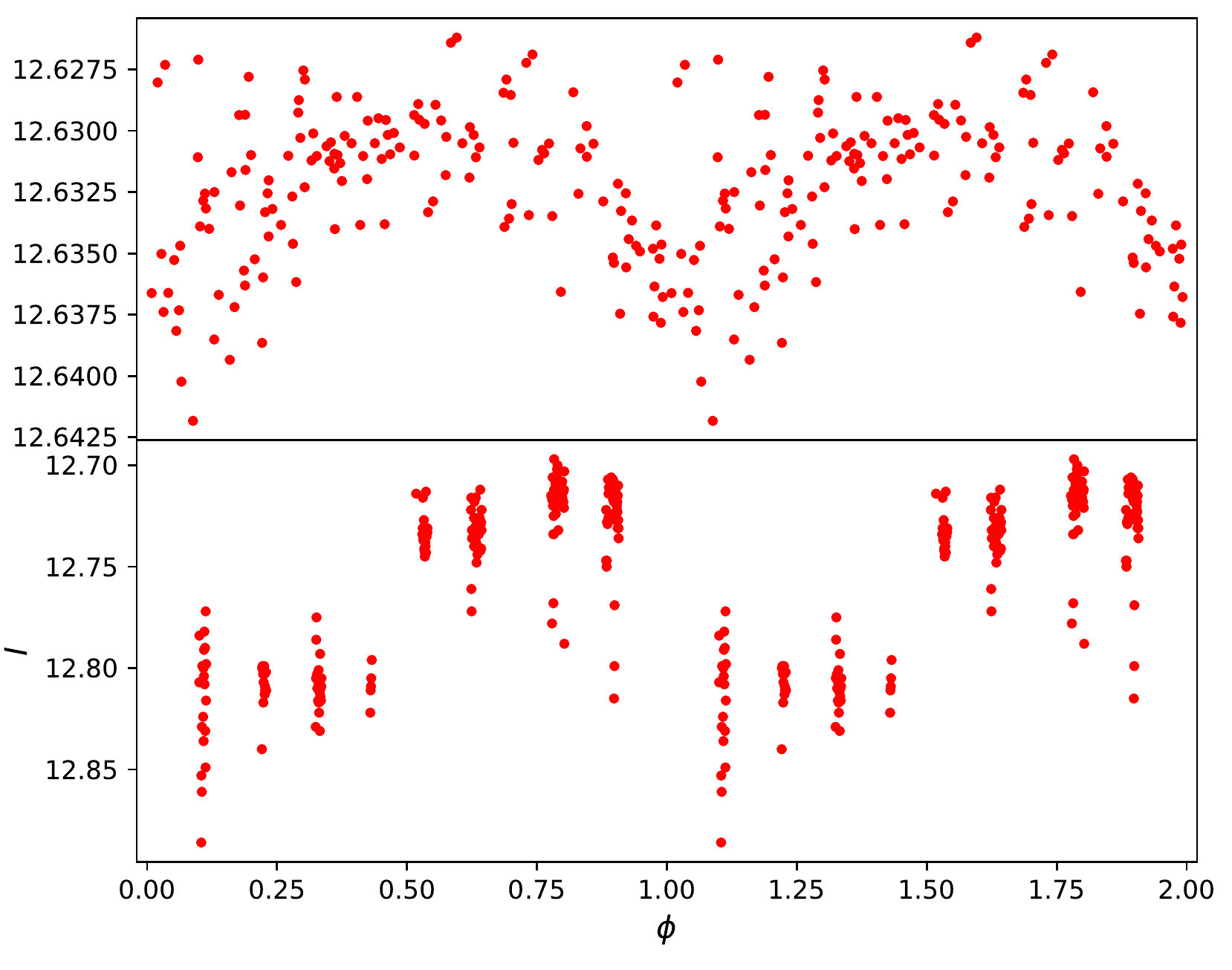}{0.33\textwidth}{(d) Star N$^{\mathrm{o}}$ 4 vs V96}
          \fig{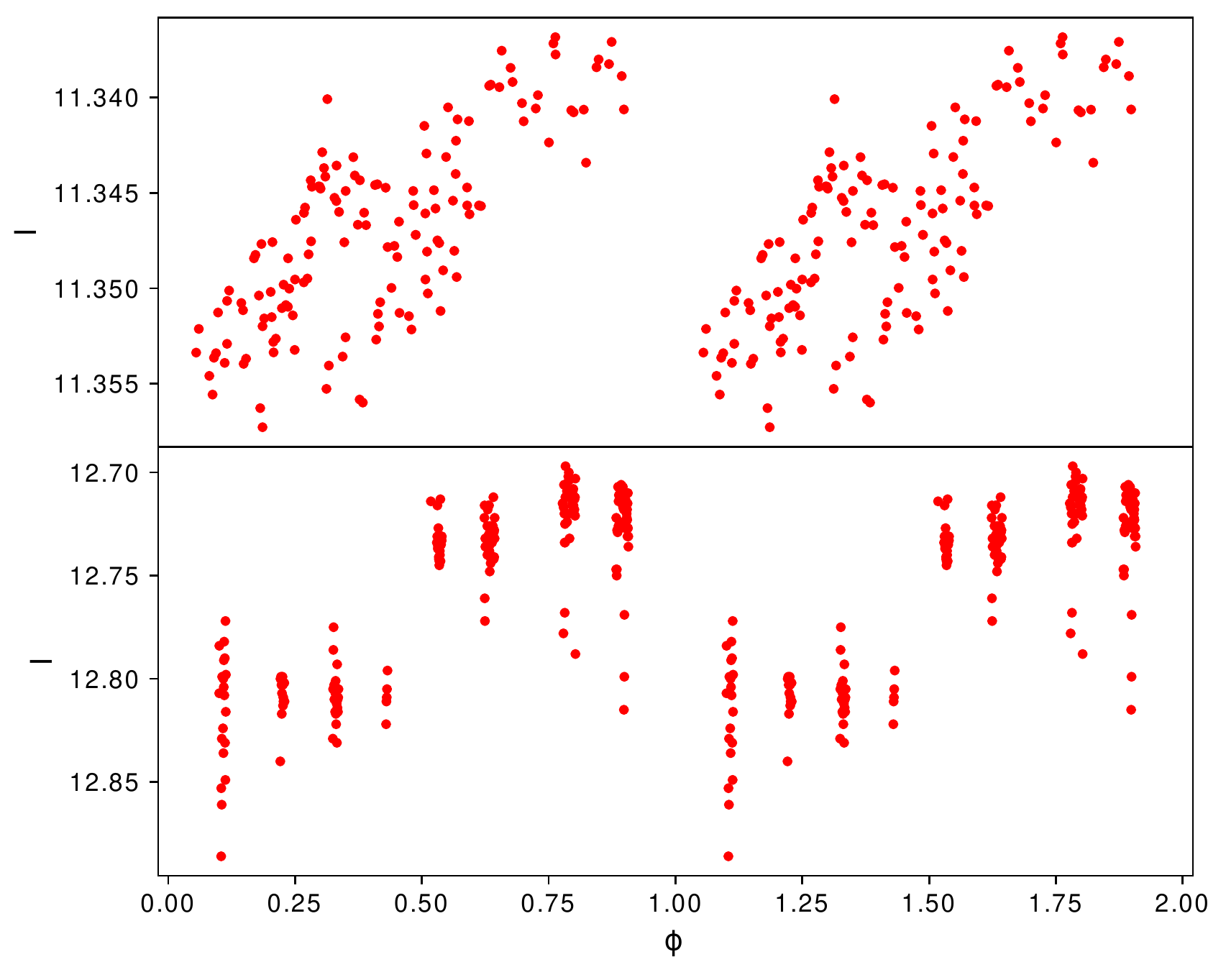}{0.33\textwidth}{(e) Star N$^{\mathrm{o}}$ 5 vs V96}
          \fig{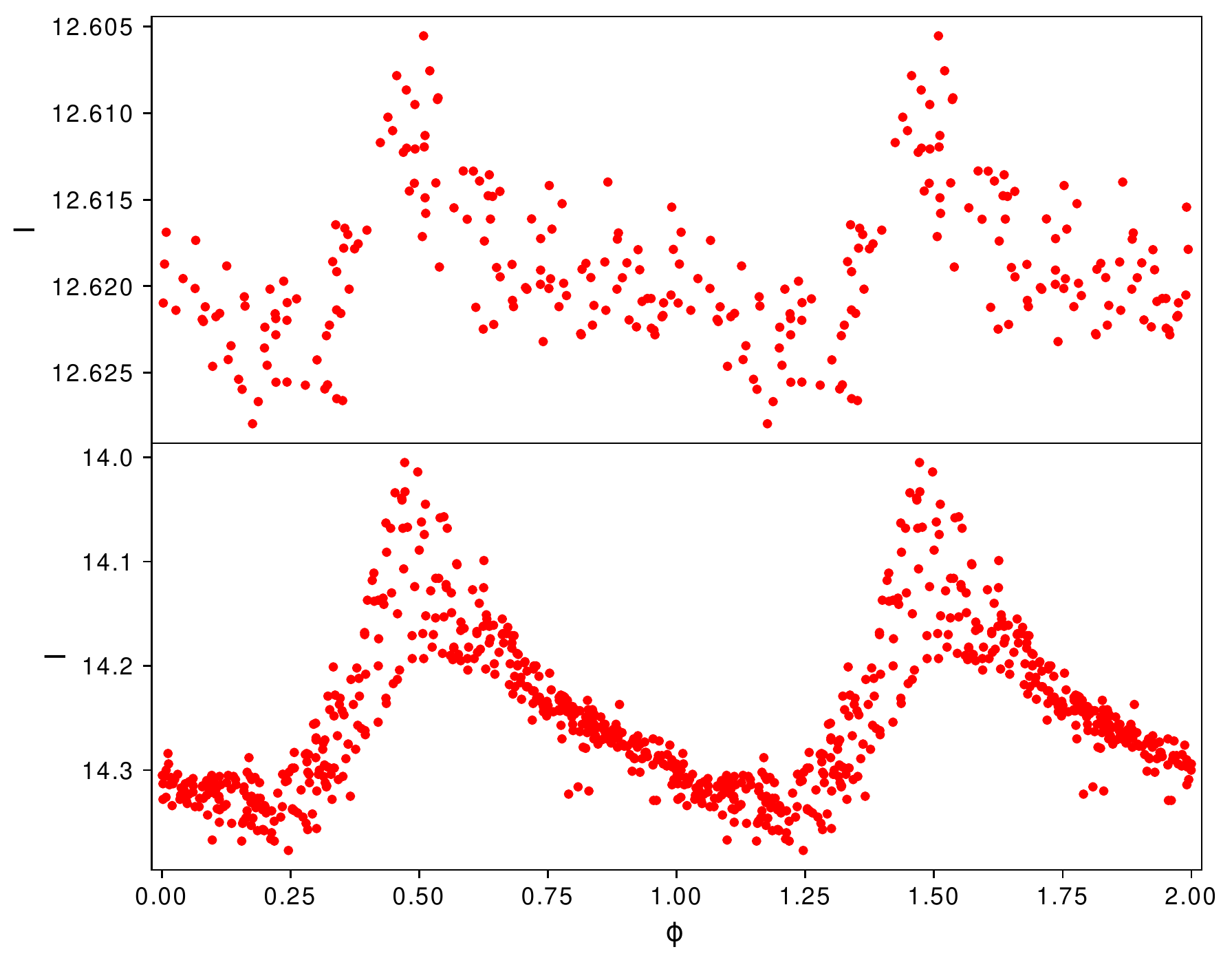}{0.33\textwidth}{(f) Star N$^{\mathrm{o}}$ 8 vs $\delta$ Scuti OGLE-LMC-DSCT-0048}
          }          
\gridline{\fig{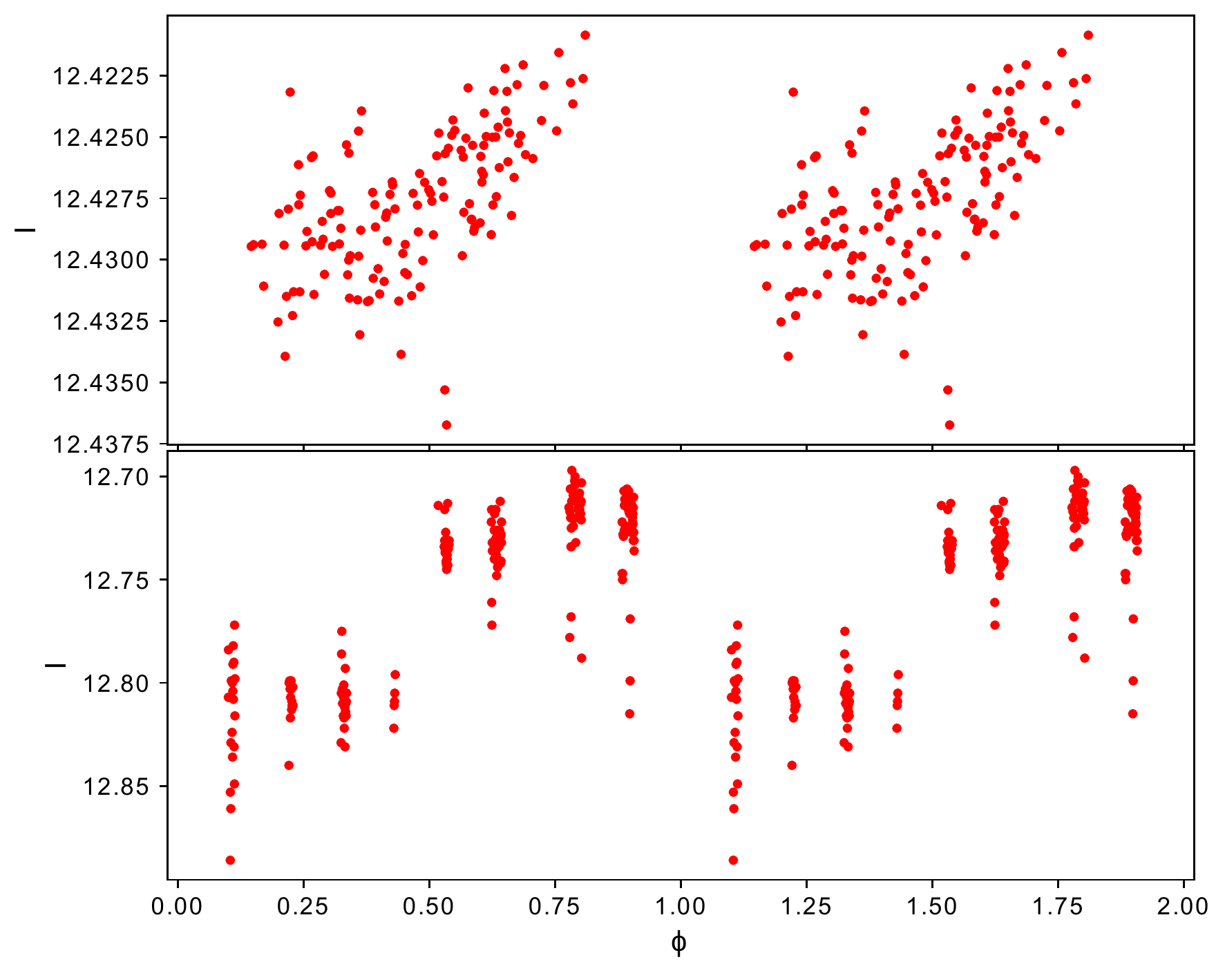}{0.33\textwidth}{(g) Star N$^{\mathrm{o}}$ 9 vs V96}
          \fig{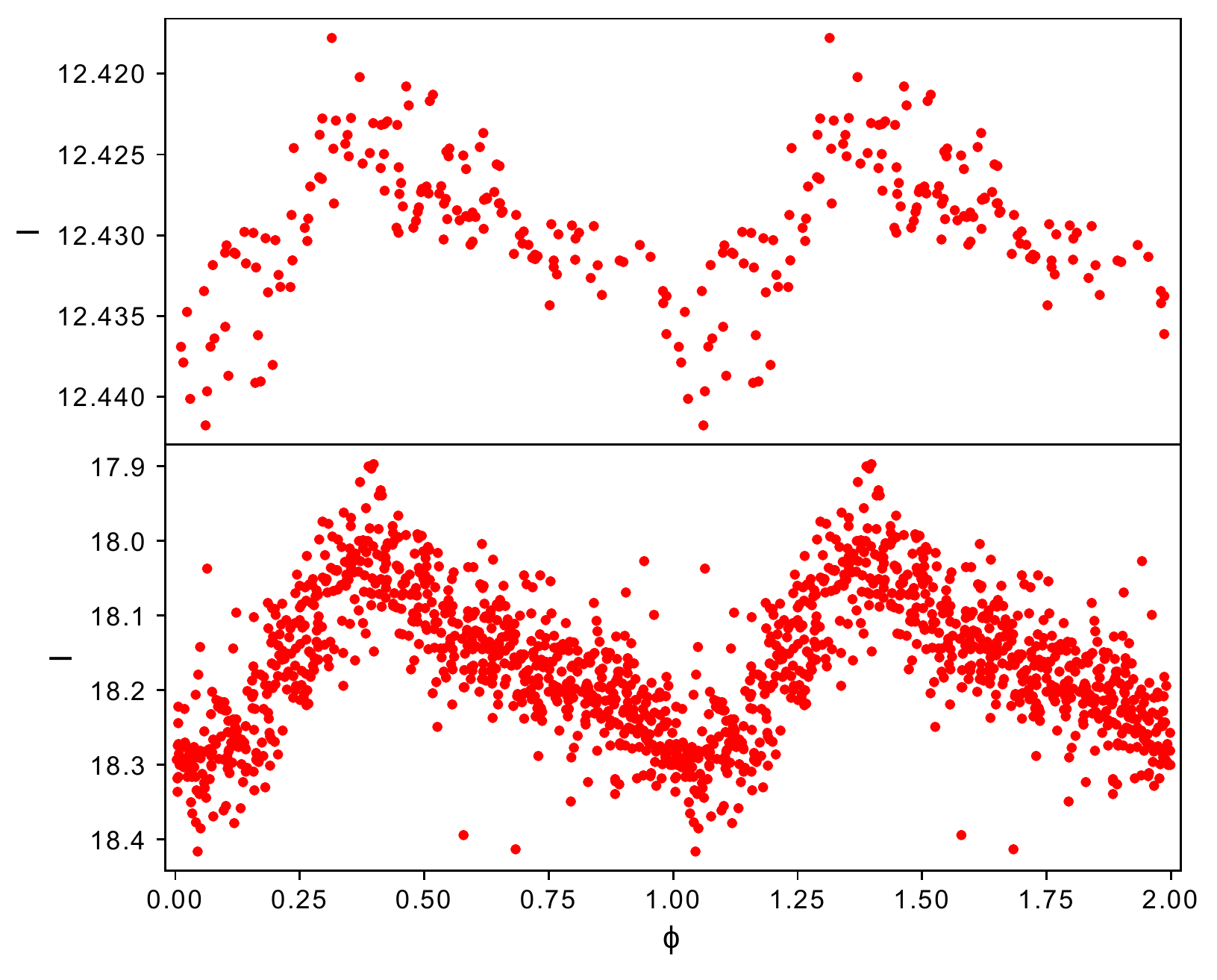}{0.33\textwidth}{(h) Star N$^{\mathrm{o}}$ 17 vs Cepheid OGLE-BLG-CEP-13}
          }          
\caption{Comparison between our sample of our variables stars(top panels) and variable stars from the OGLE catalogue of variable stars---an RR Lyrae ab, a $\delta$ Scuti, and a Cepheid variable---and the SR variable V96 from \citet{2018RMxAA..54...15Y} (bottom panels). Both comparisons are made in $I$ filter, with each star being compared with the known variable that resembles the most based on its phased light curve from Figure~\ref{fig:plotstodos} and from what was presented in Section~\ref{subsec:variables}.}
    \label{fig:Comparacion}
\end{figure*}

\textbf{Star N$^{\mathrm{o}}$ 2:} This star displays an evident magnitude difference, as shown in both LCs (Figure~\ref{fig:LCsAllV} and Figure~\ref{fig:LCsAllI}). For this case, the period analysis indicates that this star has a period $P=0.5329\pm0.0026$~days, but the phased LC shown in Figure~\ref{fig:plotstodos}\textcolor{blue}{b} does not display a complete curve. This is due to the diurnal interruption of the observations. Figure~\ref{fig:CMDcluster} shows that this star is positioned at the top of the RGB, and Figure~\ref{fig:Comparacion} shows that its phased LC resembles that of a SR variable like the V96 described by \citet{2018RMxAA..54...15Y} in the RGB of the GC NGC~6934 and are located in a similar position of the CMD near the tip. However, since SR variables display large periods, in this particular case V96 has a period of $9.54$ days, we cannot attempt to classify this star as an SR due to the discrepancy between these parameter. {We do however note that the period shown for this star is similar to those displayed in RR Lyrae stars.}

\textbf{Star N$^{\mathrm{o}}$ 3:} {The} $V$ filter LC {of this star} shows a quite small difference in magnitude, while in $I$ it is more evident, particularly during the 3rd night. From the results obtained after a more precise analysis we found a period $P=0.3527\pm0.0037$. This star is positioned in the CMD in the middle part of the RGB, and Figure~\ref{fig:plotstodos}\textcolor{blue}{c} shows the phased light curve where a small but clear variability is visible {with its period resembling those of pulsating variables (e.g., RR Lyrae stars)}. Figure~\ref{fig:Comparacion}, like in the case of Star N$^{\mathrm{o}}$ 2 shows that its phased LC is similar to the one from V96 described by \citet{2018RMxAA..54...15Y} in the RGB of the GC NGC~6934. But both the location in the CMD as well as both the period and amplitude do not agree with these kind of variables.

\textbf{Star N$^{\mathrm{o}}$ 4:} This particular star was also considered a possible variable candidate due to the significance values from Table~\ref{tab:signif}, with $S_V < 2$, but $S_I > 3$. We therefore re-performed our period analysis with more caution and detail. This star shows similar values of $P$ from the GLS method for both filters, while for the PDM method the situation was unclear. However, after restricting the P interval and taking caution that the periodogram does not display the estimated period as the only possible one, both methods yielded similar values of $P$, with the final result being {$P=0.1628\pm0.0008$}  (Table~\ref{tab:Results}). Figure~\ref{fig:CMDcluster} shows that this star is located in the lower part of the giant branch. The phased light curve of this star is in Figure~\ref{fig:plotstodos}\textcolor{blue}{d} and as in the previous star, Figure~\ref{fig:Comparacion} shows that its phased LC could resemble that of a SR variable like the V96. However, like Star N$^{\mathrm{o}}$ 3, the period shown for this star does not agree with the period of typical SR stars.  We also note that the phased LC of this star could also resemble that of an RR Lyrae type RRc displaying a similar period to those kind of variables, but we went for the SR variable due to a better resemblance of its phased LC.

\textbf{Star N$^{\mathrm{o}}$ 5:} This star shows a quite clear variable behavior as far as V filter is concerned, while for the I filter the variability is not clear (see Figure~\ref{fig:LCsAllV} and~\ref{fig:LCsAllI}). The variability analysis shows that this star have a period $P=0.4827\pm0.0024$~d; its phased light curve is reported in Figure~\ref{fig:plotstodos}\textcolor{blue}{e}. We can also see that this star, from its position in the CMD it displays a similar case a Star N$^{\mathrm{o}}$ 2, located at the top of the giant branch {and displaying a period similar to those found in RR Lyrae variables,} and Figure~\ref{fig:Comparacion} shows that its phased LC is similar to that of V96 from \citet{2018RMxAA..54...15Y} but again its period and amplitude does not agree with these type of variables.

\textbf{Star N$^{\mathrm{o}}$ 8:} This star displays a period  $P=0.0881\pm0.0001$ days, and was the one that presented the most prominent variability along with Star N$^{\mathrm{o}}$~1 although with a smaller amplitude; it was also the star on which we performed all the parameter testing from the methodology presented in \S~\ref{sec:3analisis}, where we tested how the PyAstronomy routines worked and adjusted the final parameters in order to have precise results. Similar to the case of Star N$^{\mathrm{o}}$ 1, this star's position in the CMD is in the lower part of the RGB, with its phased LC resembling more like a $\delta$ Scuti star, but with an amplitude lower than those type of stars (see Figure~\ref{fig:Comparacion}). Figure~\ref{fig:plotstodos}\textcolor{blue}{f} shows the phased light curve of this star for $V$ and $I$ filters.

\textbf{Star N$^{\mathrm{o}}$ 9:} This star displayed in its $V$ filter LC a variable behavior. The $I$ LC also displays a variability, although not as evident as in the V filter. After performing our analysis, the most probable period for the $V$ filter was $P=0.5281\pm0.0054$ days. The PDM method for the I filter suggested also another period of $P\sim0.3417$ days, that however is much less probable that the first one. From Figure~\ref{fig:CMDcluster}, we can see that this star is positioned at the lower part of the AGB {with its period being similar to those found in typical RR Lyrae stars}, and Figure~\ref{fig:Comparacion} shows that its LC resembles that of a SR variable like the V96 from \citet{2018RMxAA..54...15Y}, however as in other stars that were compared with the SR V96, its period and amplitude do not agree at all with this kind of variables. The left panel of Figure~\ref{fig:plotstodos}\textcolor{blue}{g} displays the phased light curve for this star.

\textbf{Star N$^{\mathrm{o}}$ 17:} Its LC for $V$ filter does  not display an evident variable behavior, but its LC in $I$ does show  evident variability in magnitude, which can be seen in the plot from the fifth row, first column of Figure~\ref{fig:LCsAllI}. Due to this feature and also to its significance $S_I$ value from Table~\ref{tab:signif}, which is higher that $3$, we re-performed our period analysis obtaining a period $P=0.3535\pm0.0035$~d and the amplitudes $A(V)$ and $A(I)$ shown in Table~\ref{tab:Results}. Also, based on the shape of its phased LC, like Stars N$^{\mathrm{o}}$ 1 and 8, its variable behavior is similar {to that of a pulsating variable, like a RR Lyrae or a Cepheid. In this case we chose Cepheid as the best match (see Figure~\ref{fig:Comparacion})}, but again its low amplitude value makes it difficult to provide such classification. The phased light curve of this star is shown in the right panel of Figure~\ref{fig:plotstodos}\textcolor{blue}{h}.

%%%%%%%%%%%%%%%%%%%%%%%%%%%%%%%%%%%%%%%%%%%%%%%%%%
\subsection{Possible Candidates}\label{subsec:posibles}
In this section we present the two stars that, in spite of not showing a clear variable behavior compared to the eight stars presented in the previous Section, are considered as possible variable according to the results of our period analysis and are good targets for future studies using a more complete multi-epoch photometric approach.

\begin{figure*}
\centering
	\includegraphics[width=0.9\textwidth]{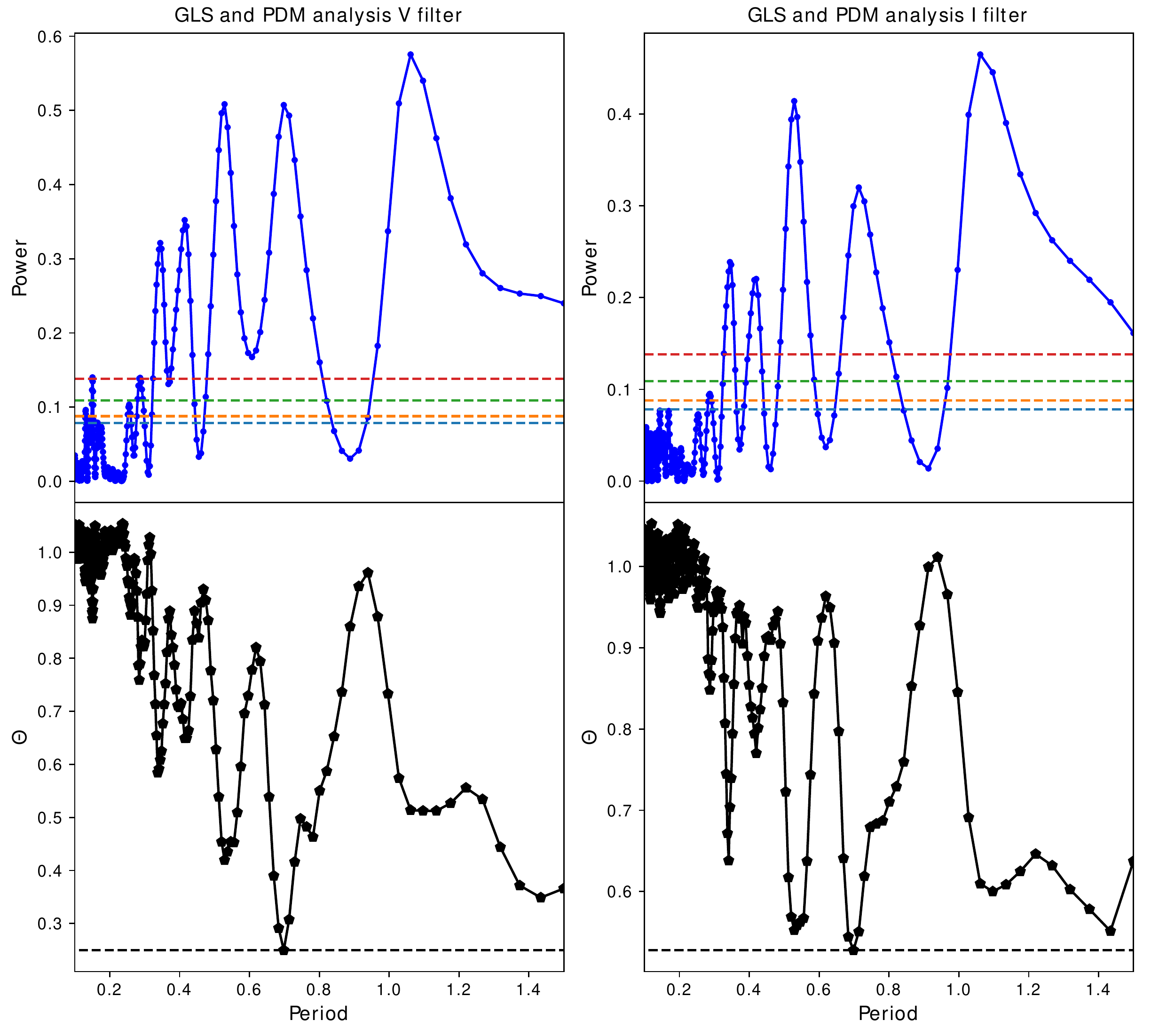}
    \caption{Resulting plots from the GLS (top panels) and PDM (bottom panels) analysis performed with the for Star N$^{\mathrm{o}}$ 13. Note that the GLS analysis marks the highest period $P=1.0614$~d, while the PDM analysis mark a period $P=0.6983$~d for both filters, meanwhile the second highest period for the GLS analysis in both filters correspond to a period $P=0.5287$.}
    \label{fig:PlotsStar13Pend15}
\end{figure*}

\begin{figure*}
\gridline{\fig{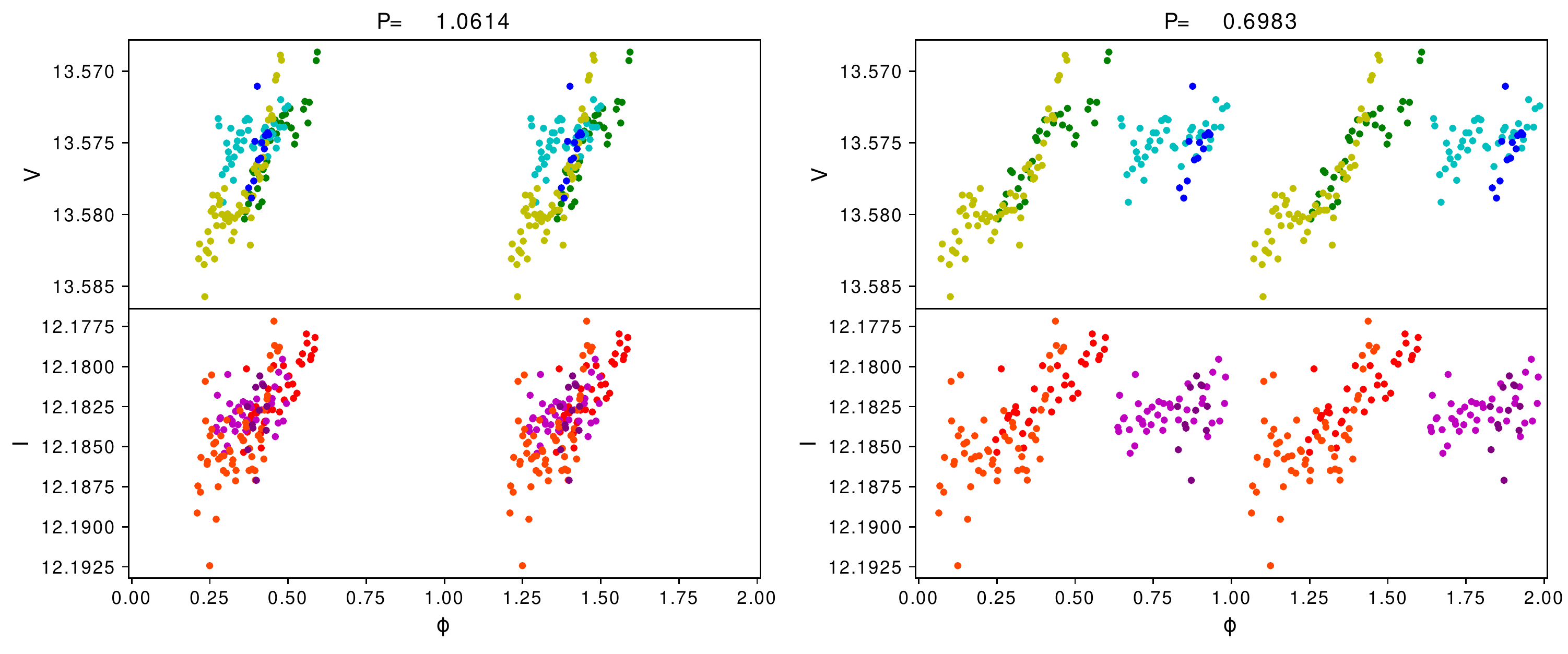}{\textwidth}{}
}
\gridline{\fig{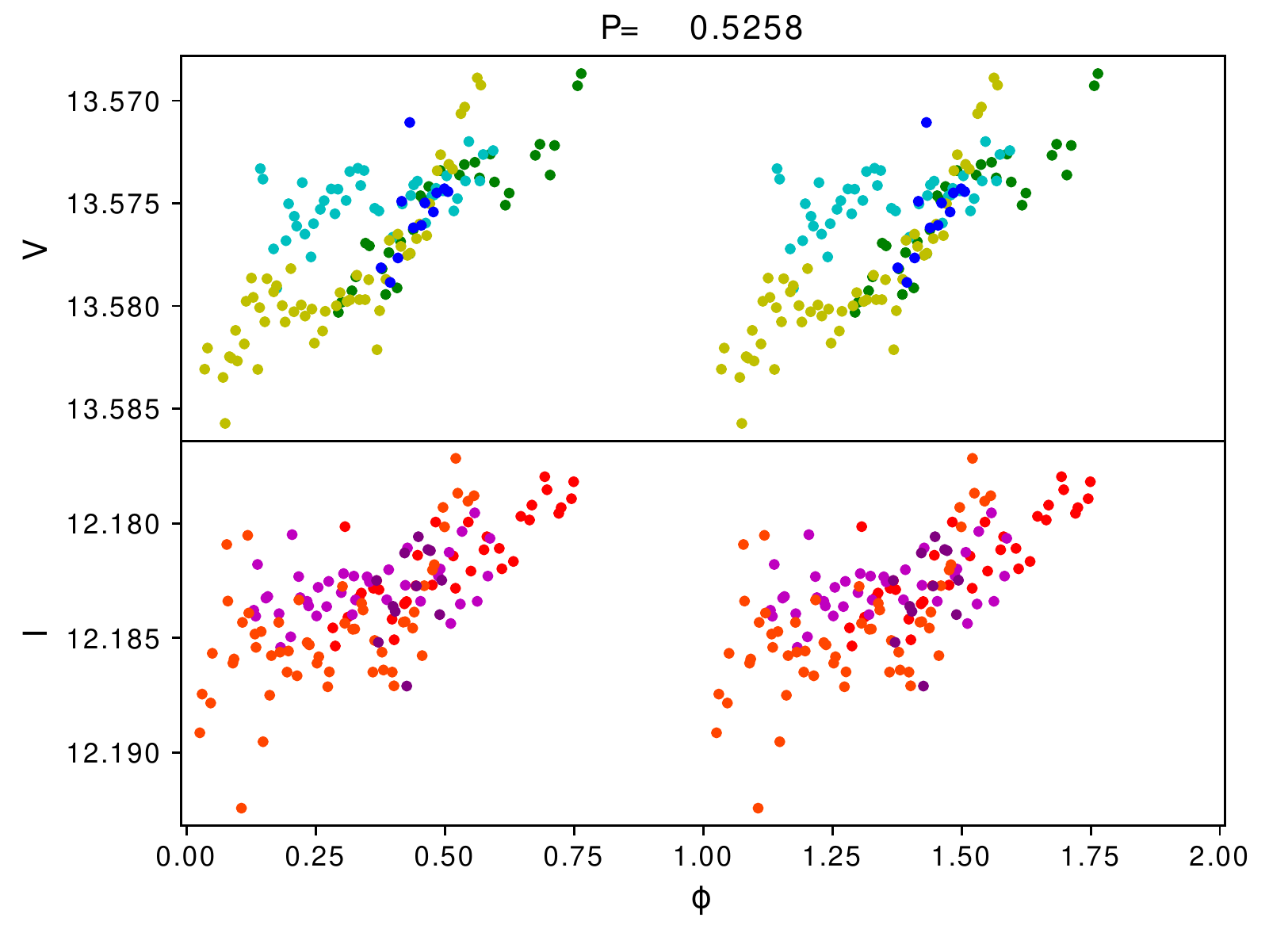}{0.5\textwidth}{}
}
\caption{Different phased light curves of Star N$^{\mathrm{o}}$ 13, with each plot displaying a different possible period period, with top left being $P=1.0614$~d, top right $P=0.6983$~d and the bottom panel $P=0.5287$. Once again, the different colors are the same as in Figure~\ref{fig:plotstodos}}
\label{fig:FasePosible13}
\end{figure*}

\textbf{Star N$^{\mathrm{o}}$ 13:} This possible variable star is interesting since both LCs displayed a variable behavior during the 3rd night (although quite small). The GLS method derived a period $P=1.0614$~d but the PDM analysis showed a period $P=0.6983$~d, even thought the second highest probability period for the GLS analysis was at $P=0.5287$ (top panels of Figure~\ref{fig:PlotsStar13Pend15}). We adjusted the PyAstronomy parameters to evaluate the other periods. It is worth to note that there are two more periods above the 0.1\% threshold at $P\sim0.41564$ and $P\sim0.33818$, but the amplitudes derived by the GLS analysis were equal (if not lower) than the mean magnitude error, so they were rejected. Figure~\ref{fig:FasePosible13} shows the phased light curves of the three different periods. Future observations for this star could help to determine if this star is a real variable and, if so, its variability period.

\begin{figure*}
	\includegraphics[width=\textwidth]{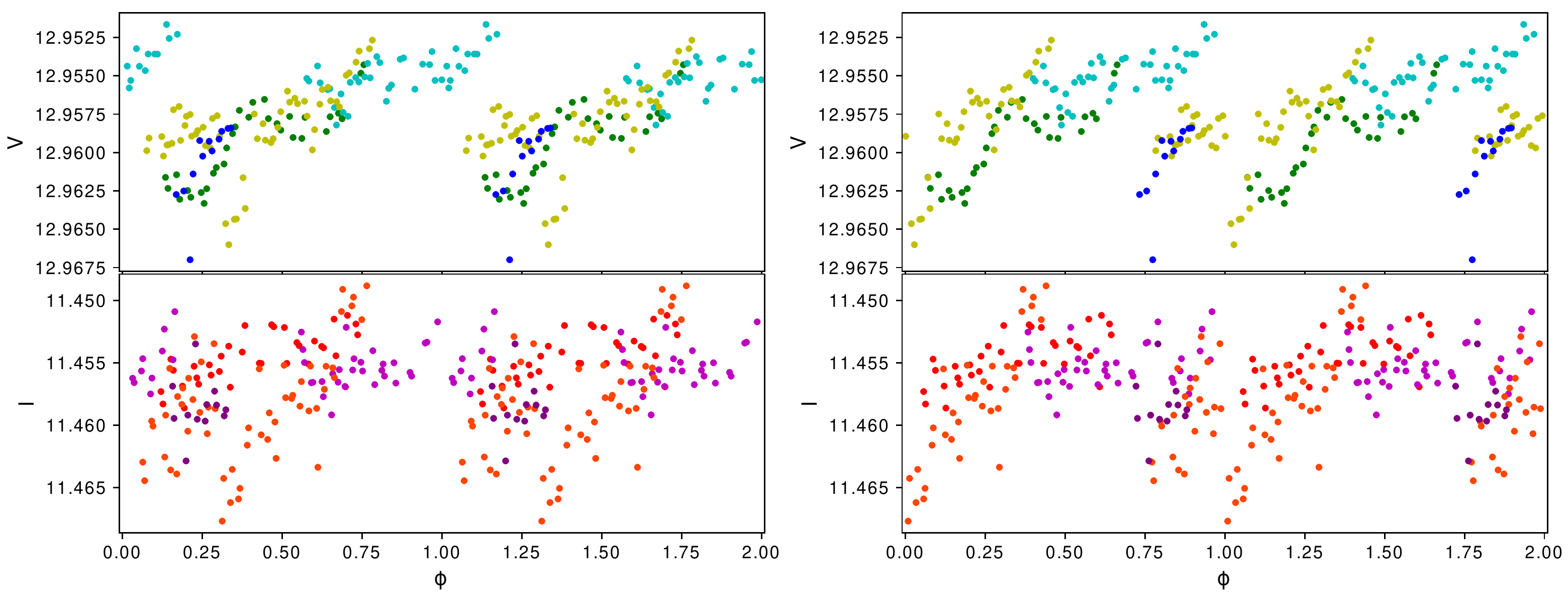}
    \caption{Left panels: Phased Light Curve of Star N$^{\mathrm{o}}$ 16 for $V$ (top panel) and $I$ (bottom panel) filters, plotted by phasing the light curve with the period obtained from our analysis in the $V$ filter $P_V=0.3981$ days. As in previous figures, the different colors are the same as in Figure~\ref{fig:plotstodos}. Right panels: Phased Light Curve of Star N$^{\mathrm{o}}$ 16 for $V$ (top panel) and $I$ (bottom panel) filters, plotted by phasing the light curve with the period obtained from our analysis in the in the $I$ filter $P_I=0.4174$ days. As in previous figures, the different colors are the same as in Figure~\ref{fig:plotstodos}.}
    \label{fig:fases4ambas}
\end{figure*}

\textbf{Star N$^{\mathrm{o}}$ 16:} This particular star was challenging since its LC for $V$ filter did not present any magnitude {variation}, but for the $I$ filter it displayed a peculiar behavior during the 3rd night (4th row and 4th column panel of Figure~\ref{fig:LCsAllI}). We derived {two different periods for $V$ and $I$ filters} ($P_V=0.3981$~d and $P_I=0.4174$~d, left and right panels of Figure~\ref{fig:fases4ambas}, respectively), as in the case of Star  N$^{\mathrm{o}}$ 13, future observations for could help to determine if this stars is a real variable and, if so, its variability period.

If the variability behaviour of star 13 and 16 is confirmed, a classification could be possible for both stars in a similar manner as in the case of our variable stars from \S~\ref{subsec:variables}.

%%%%%%%%%%%%%%%%%%%%%%%%%%%%%%%%%%%%%%%%%%%%%%%%%%
\subsection{Variability and its possible impact on the spectroscopic [Fe/H] determination}\label{subsec:variymetalicidad}

After finishing our period analysis of the variable stars, the next step was to check if their variable behavior has an impact on the $\text{[Fe/H]}$ determination as obtained from spectra. First of all, we used the abundance study of NGC~3201 by \cite{2015ApJ...801...69M}, who measured the iron abundance by adopting photometric values of $\log g$. They used photometric gravities  based on the results of \citet{2014ApJ...797..124L}, who suggested that NLTE effects could affect in AGB stars the abundances that are obtained from Fe~I lines, although the abundances from Fe~II lines appear not to be modified. This effect can be detected by assuming photometric values of $\log g$ and measuring the abundances from Fe~I and Fe~II independently. They also remarked that RGB stars do not display such effects. We point out here that AGB and RGB stars in the sample have very similar atmospheric parameter. For this reason, if they are affected by NLTE, the final Fe~I abundances should suffer the same change, independently of the evolutionary stage.

\begin{figure*}
	\includegraphics[width=\textwidth]{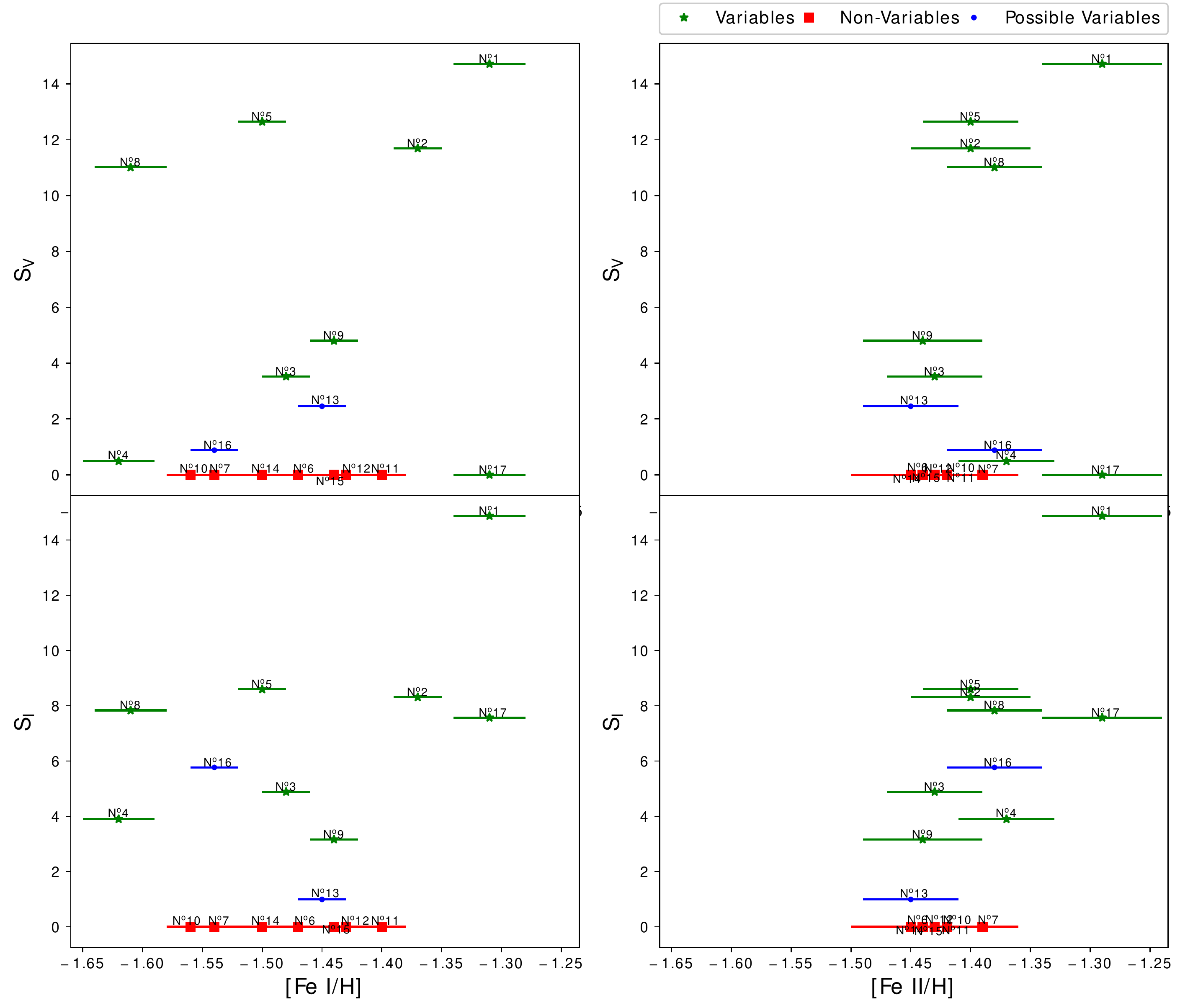}
    \caption{{Left panels: Significance $S_V$ and $S_I$ values for each star plotted against their respective $\text{[Fe I/H]}$ measurements from \citet{2015ApJ...801...69M}. The green stars represent the variable stars, the red squares represent the non-variables and the blue dots represent the possible candidate variables. Right panels: Significance values for each star plotted against their $\text{[Fe II/H]}$ measurements from \citet{2015ApJ...801...69M} plotted with the same x-axis scale as the left panel and the different colors for each group of stars are the same as in the left panels.}}
    \label{fig:histoFeIyFeII}
\end{figure*}

We grouped each star according to their classification given in \S~\ref{sec:3analisis} and \S~\ref{sec:resultados}---variable, non-variable and possible variable---and, with the [Fe I/H] values shown in Table~\ref{tab:datosstars}, we plotted the significance $S_V$ (top panels) and $S_I$ (bottom panels) against their [Fe I/H] for each group in the left panel of Figure~\ref{fig:histoFeIyFeII}. In this figure we can see that: stars that have extreme [Fe I/H] values are variables, i.e., the most metal-poor stars like Star N$^{\mathrm{o}}$~8 ($\text{[Fe I/H]} = -1.61\pm0.03$) and N$^{\mathrm{o}}$~4 ($\text{[Fe I/H]} = -1.62\pm0.03$), and the most metal-rich like stars N$^{\mathrm{o}}$~1 and N$^{\mathrm{o}}$~17 with $\text{[Fe I/H]} = -1.31\pm0.03$, and Star N$^{\mathrm{o}}$~2 with $\text{[Fe I/H]} = -1.37\pm0.02$ are located at the extremes of the [Fe I/H] distribution. However, the 3 remaining variable, i.e. stars N$^{\mathrm{o}}$~3, 5, and 9, have a metallicity that well agrees with that of the non-variable sample. This could indicate, as we will infer in the discussion from the study of \citet{2011ApJS..197...29F}, that the [Fe I/H] metallicity as obtained from the spectra could depend on the phase at which the spectrum is taken. Furthermore, we note that stars that are not variable, and those which are possible variables (\S~\ref{subsec:posibles}) have [Fe I/H] values that are close to the mean value of [Fe I/H] of the total sample.

We also checked the $\text{[Fe II/H]}$ content of each variable star to see if it shows a similar behavior. The right panels of Figure~\ref{fig:histoFeIyFeII} shows the significance values of the variable stars, possible variables and non-variables plotted against [Fe II/H]. From this figure we can see that variable stars tend to have on average higher [Fe II/H] values compared with non-variable and as that, as in the case of [Fe I/H], variable stars show a higher spread degree in comparison with the possible and non-variables.

\begin{figure}
\centering
	\includegraphics[width=0.7\textwidth]{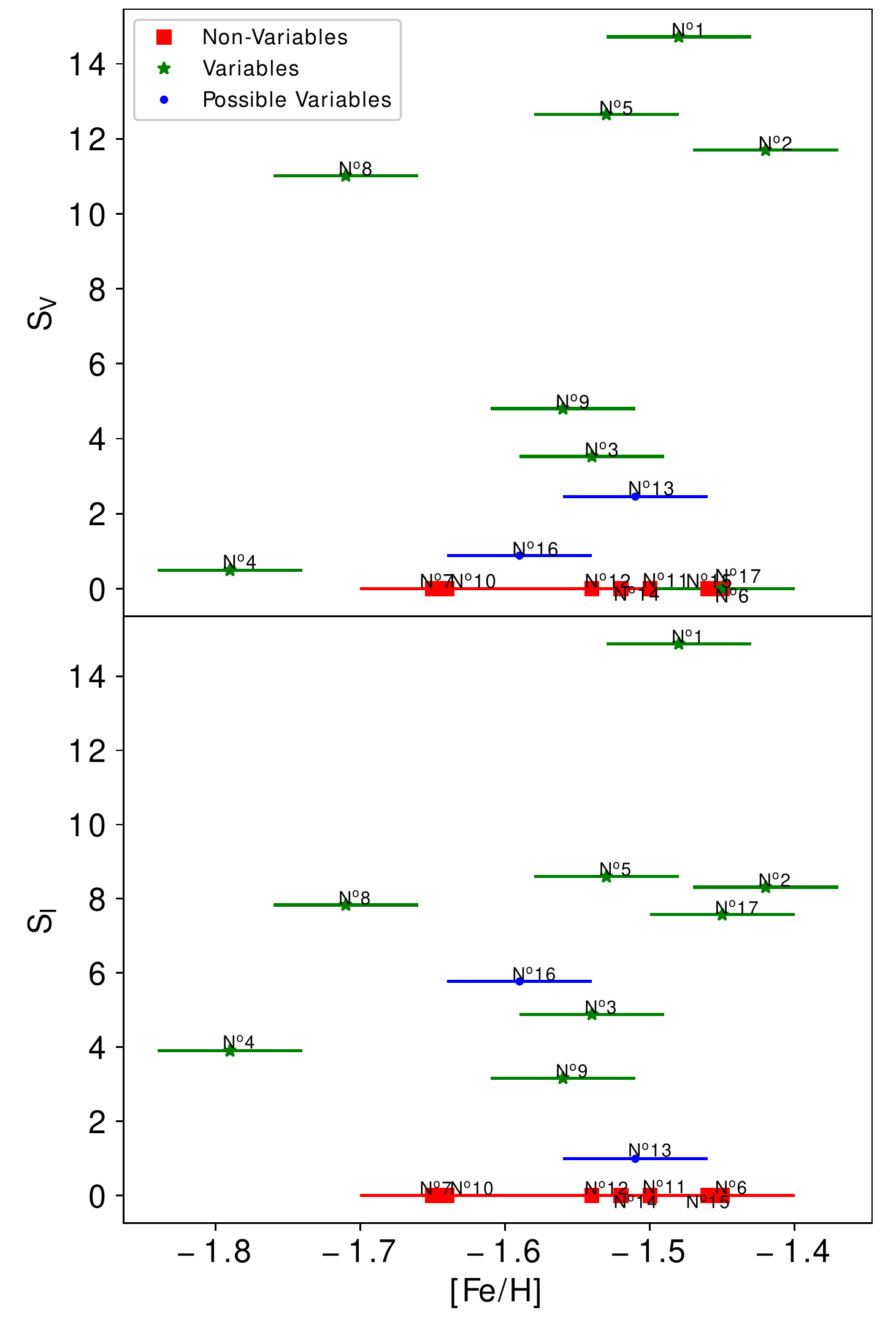}
    \caption{{Top panel: plot displaying the significance $S_V$ values of the 17 stars against their $\text{[Fe/H]}$ abundance from \citet{2013ApJ...764L...7S}. Bottom panel: Significance $S_I$ values of the 17 giants plotted against their $\text{[Fe/H]}$. The colors for both panels are the same as in Figure~\ref{fig:histoFeIyFeII}. For the error bar we assumed an error on [Fe/H] of 0.05 dex, that is the typical error for purely spectroscopic measurement.}}
    \label{fig:histoFeSimm2013}
\end{figure}

As a second step we used the abundance from \citet{2013ApJ...764L...7S} to see if also in this case we obtain the same result. We point out the fact that \citet{2013ApJ...764L...7S} uses spectroscopic parameters and for this reason they have $\text{[Fe I/H]}=\text{[Fe II/H]}$ and so a unique value of iron ([Fe/H]) for each star. Figure~\ref{fig:histoFeSimm2013}, shows the significance $S_V$ (top panel) and $S_I$ (bottom panel) of our 17 giants plotted against their $\text{[Fe/H]}$. From this plot we can see again that stars with extreme values of $\text{[Fe/H]}$ (both metal-poor and metal-rich) are variable. However in this case the most metal-poor variable are those that have a larger [Fe/H] difference with respect the non-variable stars.

Table~\ref{tab:TablaAbundancias} shows the final classification of all 17 stars along with the iron abundance measurements from \citet{2013ApJ...764L...7S} and \citet{2015ApJ...801...69M} and their significance $S_V$ and $S_I$. Note that for stars that were classified as non-variable both significance values $S_V$ and $S_I$ were assumed to be zero. If we move to statistics, it is worth to report the mean abundances and spreads with their errors of the three groups of stars calculated for [Fe I/H], [Fe II/H] and [Fe/H]. These values can be found in Table~\ref{tab:TablaSigmasFe}. We consider the mean abundances first. We see that mean [Fe I/H] from \citet{2015ApJ...801...69M} agrees well for the three groups, especially for the variable and the non-variable. For mean [Fe II/H] instead, variable stars show a value that is 0.05 higher than the non-variables, beyond the statistics errors. As far as \citet{2013ApJ...764L...7S} is concerned, the agreement is well within $1\sigma$ for all three groups.

The effect of variability appears when we consider the iron spread. For \citet{2015ApJ...801...69M}, [Fe I/H] spread for possible and non-variables is of $0.06$ dex, while the spread for variables is much larger with a value of $0.12$ dex. The same is true if we compare [Fe II/H] spreads for variable and non-variables in \citet{2015ApJ...801...69M} and [Fe/H] spreads for the same groups in \citet{2013ApJ...764L...7S}.

We conclude the variability has a negligible effect on mean iron metallicities with a possible but small increase of mean [Fe II/H] compared with [Fe I/H] when atmospheric parameters are obtained from photometry as in \citet{2015ApJ...801...69M}. On the other hand the effect on the iron spread is important. We see in fact that variable stars have a much higher iron spread regardless the method used to obtain the parameters. The spread for variable is up to twice the values obtained for non-variable. This effect must be taken into account in any study of intrinsic iron spread of stellar populations, especially in globular clusters, otherwise the wrong conclusion could be that the cluster hosts stars with different metallicities.

\begin{table}
	\centering
	\caption{Final classification, iron abundance information and significance values of the 17 giant stars.}
	\label{tab:TablaAbundancias}
	\begin{tabular}{lcccccc} 
		\hline
		Star & Classification & $\text{[Fe/H]}$ & $\text{[Fe I/H]}$  & $\text{[Fe II/H]}$ & {$S_V$} & {$S_I$} \\
		 & & (dex) & (dex) & (dex) &  &  \\
		\hline
		1 &Variable&$-1.48$& $-1.31\pm0.03$&$-1.29\pm0.05$ &{$14.719$} &{$14.874$}\\
		2 &Variable&$-1.42$& $-1.37\pm0.02$&$-1.40\pm0.05$ &{$11.695$} &{$8.311$}\\
		3 &{Variable}&$-1.54$& $-1.48\pm0.02$&$-1.43\pm0.04$ &{$3.524$} &{$4.885$}\\
		4 &Variable&$-1.79$& $-1.62\pm0.03$&$-1.37\pm0.04$ &{$0.490$} &{$3.900$}\\
		5 &Variable&$-1.53$&$-1.50\pm0.02$&$-1.40\pm0.04$ &{$12.648$} &{$8.596$}\\
		6 &Non-Variable&$-1.45$& $-1.47\pm0.02$&$-1.45\pm0.05$ &{$0.0$} &{$0.0$}\\
		7 &Non-Variable&$-1.65$& $-1.54\pm0.02$&$-1.39\pm0.03$ &{$0.0$} &{$0.0$}\\
		8 &Variable&$-1.71$& $-1.61\pm0.03$&$-1.38\pm0.04$ &{$11.013$} &{$7.833$}\\
		9 &{Variable}&$-1.56$& $-1.44\pm0.02$&$-1.44\pm0.05$ &{$4.800$} &{$3.162$}\\
		10 &Non-Variable&$-1.64$& $-1.56\pm0.02$&$-1.42\pm0.04$ &{$0.0$} &{$0.0$}\\
		11 &Non-Variable&$-1.50$& $-1.40\pm0.02$&$-1.42\pm0.05$ &{$0.0$} &{$0.0$}\\
		12 &Non-Variable&$-1.54$& $-1.43\pm0.03$&$-1.43\pm0.04$ &{$0.0$} &{$0.0$}\\
		13 &{Possible}&$-1.51$& $-1.45\pm0.02$&$-1.45\pm0.04$ &{$2.454$} &{$0.990$}\\
		14 &Non-Variable&$-1.52$& $-1.50\pm0.02$&$-1.45\pm0.04$ &{$0.0$} &{$0.0$}\\
		15 &Non-Variable&$-1.46$ & $-1.44\pm0.03$&$-1.44\pm0.06$ &{$0.0$} &{$0.0$}\\
		16 &Possible&$-1.59$& $-1.54\pm0.02$&$-1.38\pm0.04$ &{$0.880$} &{$5.769$}\\
		17 &Variable&$-1.45$& $-1.31\pm0.03$&$-1.29\pm0.05$ &{$0.0$} &{$7.570$}\\
		\hline
	\end{tabular}
	\tablecomments{The $[\text{Fe/H}]$ values are from \citet{2013ApJ...764L...7S}, while the $[\text{Fe I/H}]$ and $[\text{Fe II/H}]$ values are from \citet{2015ApJ...801...69M}.}
\end{table}

\begin{table}
	\centering
	\caption{Mean, $\sigma$ and $err_{\sigma}$ of the iron abundances for each group of stars.}
	\label{tab:TablaSigmasFe}
	\begin{tabular}{lccccc} 
		\hline
		 & Classification & $\bar{X}$ & $\sigma_{\bar{X}}$ & $\sigma$&$err_{\sigma}$\\
		 & &(dex) & (dex) & (dex) & (dex)\\
		\hline
		$\text{[Fe I/H]}_{\text{Mucciarelli}}$&Variable&$-1.46$&$0.04$ &$0.12$&$0.03$\\
		 &Possible&$-1.50$&$0.03$ &$0.05$&$0.02$\\
		 & Non-Variable &$-1.48$&$0.02$ &$0.06$&$0.01$\\ 
		\hline
		$\text{[Fe II/H]}_{\text{Mucciarelli}}$&Variable&$-1.37$&$0.02$ &$0.06$&$0.01$\\
		 &Possible&$-1.41$&$0.03$ &$0.04$&$0.02$\\
		 & Non-Variable &$-1.43$&$0.01$ &$0.02$&$0.01$\\ 
		\hline
		$\text{[Fe/H]}_{\text{Simmerer}}$&Variable&$-1.56$&$0.04$ &$0.13$&$0.03$\\
		 &Possible&$-1.55$&$0.04$ &$0.04$&$0.02$\\
		 & Non-Variable &$-1.53$&$0.03$ &$0.08$&$0.02$\\ 
		\hline
	\end{tabular}
\end{table}

%%%%%%%%%%%%%%%%%%%%%%%%%%%%%%%%%%%%%%%%%%%%%%%%%%
%%%%%%%%%%%%%%%%%%%%%%%%%%%%%%%%%%%%%%%%%%%%%%%%%%
\section{Discussion}\label{sec:discuss}

At this point it is interesting to check where the different groups of stars lie in the CMD of Figure~\ref{fig:CMDcluster}. We see there that variable are distributed evenly among RGB and AGB stars, with 3 objects being AGBs and 4 being RGBs. On the other hand only 2 non-variable appear to be AGBs against 5 RGBs.  In any case we cannot reach any firm conclusion about a possible relation between variability and the evolutionary phase due to the limited sample we have.

About the possible mechanism responsible of the variability, we note that the LC of 3 of our targets (1, 8, and 17) resembles closely stars whose variability is caused by radial or no-radial pulsations (RRLyrae, Delta-Scuti, and Cepheid respectively). From that we deduce that pulsation is very likely the cause of the variability of our stars and that it is related with the spread in the spectroscopic [Fe/H] determination we observe.

\begin{figure*}
    \centering
    \includegraphics[width=0.75\textwidth]{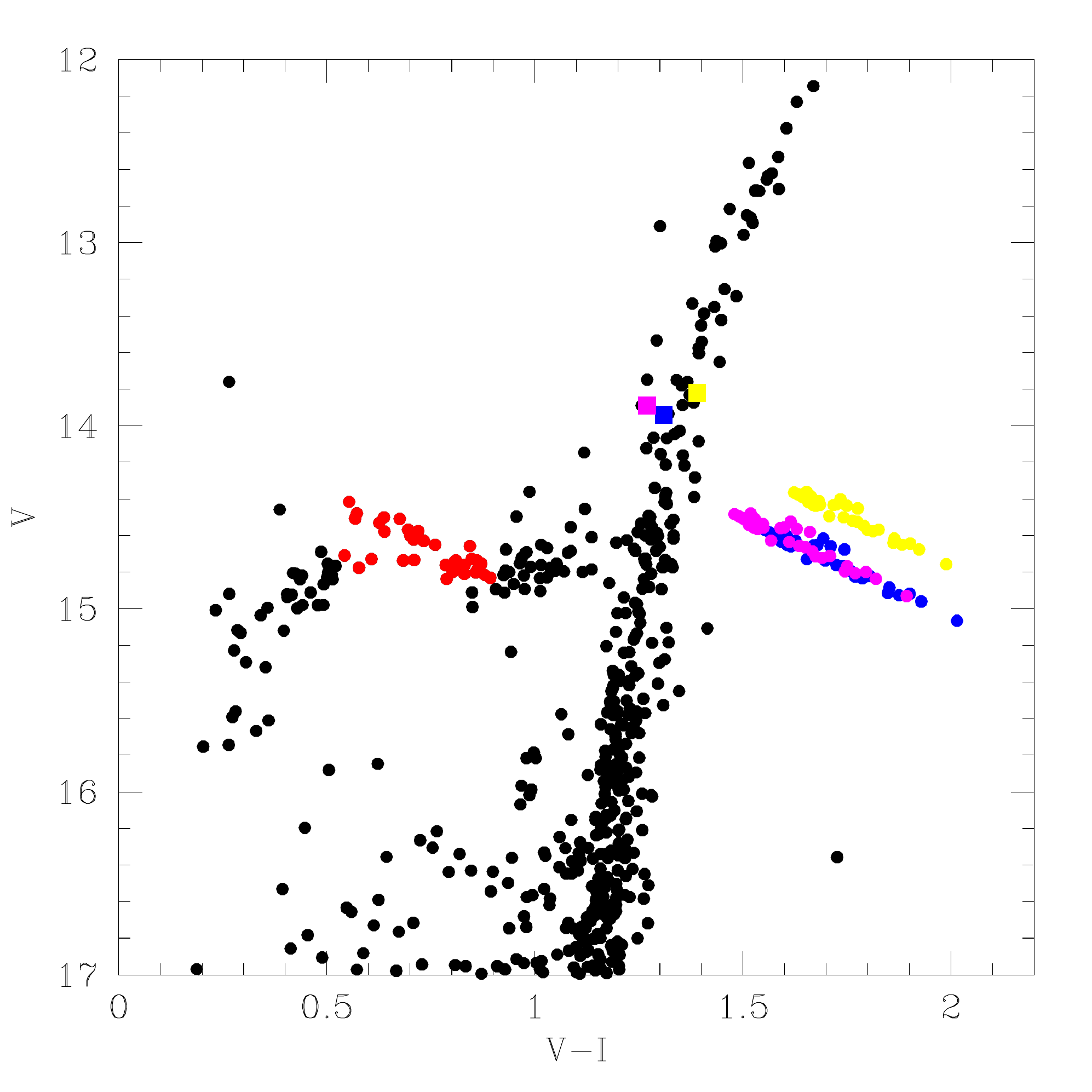}
    \caption{{CMD of NGC 3201 showing how photometric blending between a giant star and a variable star could affect in the position of stars 1, 8, and 17 from this work. Red stars are all known RR Lyrae from the cluster, while our stars are marked in different colors: Star N$^{\mathrm{o}}$1 is blue, N$^{\mathrm{o}}$ 8 is magenta and N$^{\mathrm{o}}$ 17 is yellow.}}
    \label{fig:CMDblend}
\end{figure*}

{At this point, it is also worth mentioning the possibility that the variability we observe is not intrinsic to the RGB or AGB stars we are studying, but it is due to an unresolved blending with other variables of the cluster. In order to check this hypothesis we will analyze here the three stars \#1, \#8, and \#17 since they show a LC very similar to that of pulsating variables (e.g., RR-Lyrae). Because of this and since the only known variables bright enough to affect the $V$ and $I$ magnitude of a giant stars in NGC~3201 are RR-Lyrae \citep{2014RMxAA..50..307A}, we will consider only these object as possible polluters of the LC. Figure~\ref{fig:CMDblend} shows the position in the CMD of our three stars (squares) and of the known RR-Lyrae (red circles). The aim is know what the real position of the three targets would be if we remove the contamination form the RR-Lyrae.}

{If we have two blended stars (stars $A$ and $B$) the equation to calculate the magnitude of $A$ using the total magnitude ($A+B$) and the magnitude of $B$  (in our case a RR Lyrae) can be obtained from Pogson's equation and is:}
{$${m_A = m_B - 2.5\log(10^{(-(m_{tot}-m_B)/2.5)}-1)}$$}

{We applied this equation to the $V$ and $I$ magnitudes of the three stars using the $V$ and $I$ mean magnitudes of all the RR-Lyrae. The result is the three strips of circles in Figure~\ref{fig:CMDblend}. As we can see, if the hypothesis of the blending is true, the three targets would fall well out of the RGB of the cluster. They could still be giants, but much more metal-rich than the value obtained from spectroscopy ([Fe/H]$< -1.4$). They would also fall in a region completely empty of stars and they will not be compatible with being members of the clusters at odd with what found in all the spectroscopic studies we mention in previous sections.}

{Another test we can do is to calculate the probability to find a blend between a giant stars and a RR-Lyrae. In order to do that we first have to define the area around a target where, if a RR-Lyrae falls, we have a unresolved blend. Since the Full Width at Half Maximum (FWHM) of the observation was around $2$ arcsec, it is safe to assume a circular area with a radius of the same size. Since the field of view (FOV) is 5.4'x5.4' (see Figure~\ref{fig:Area}) and that in Figure~\ref{fig:CMDblend} we identify 35 RR-Lyrae, we find that that probability of a blend is: $$P=\frac{Area_{\text{circular area}}}{Area_{\text{FOV}}}\cdot N_{\text{RR-Lyrae}}=0.004.$$}

{If we calculate then the probability to have three stars affected by a blend, we end with total a probability of: $$P_{TOT}=P^3=0.00000007$$}

{These numbers can vary if we change our inputs such as the FOV, the size of the circular area, or if we consider that the RR-Lyrae are not uniformly distributed in the FOV but they follow a King density profile. In any case the probability we calculated is low enough to assure that, together with the CMD position test, none of the LCs of our targets is the result of a blend between a regular RGB or AGB and a variable star.}

\begin{figure}
\centering
	\includegraphics[width=\textwidth]{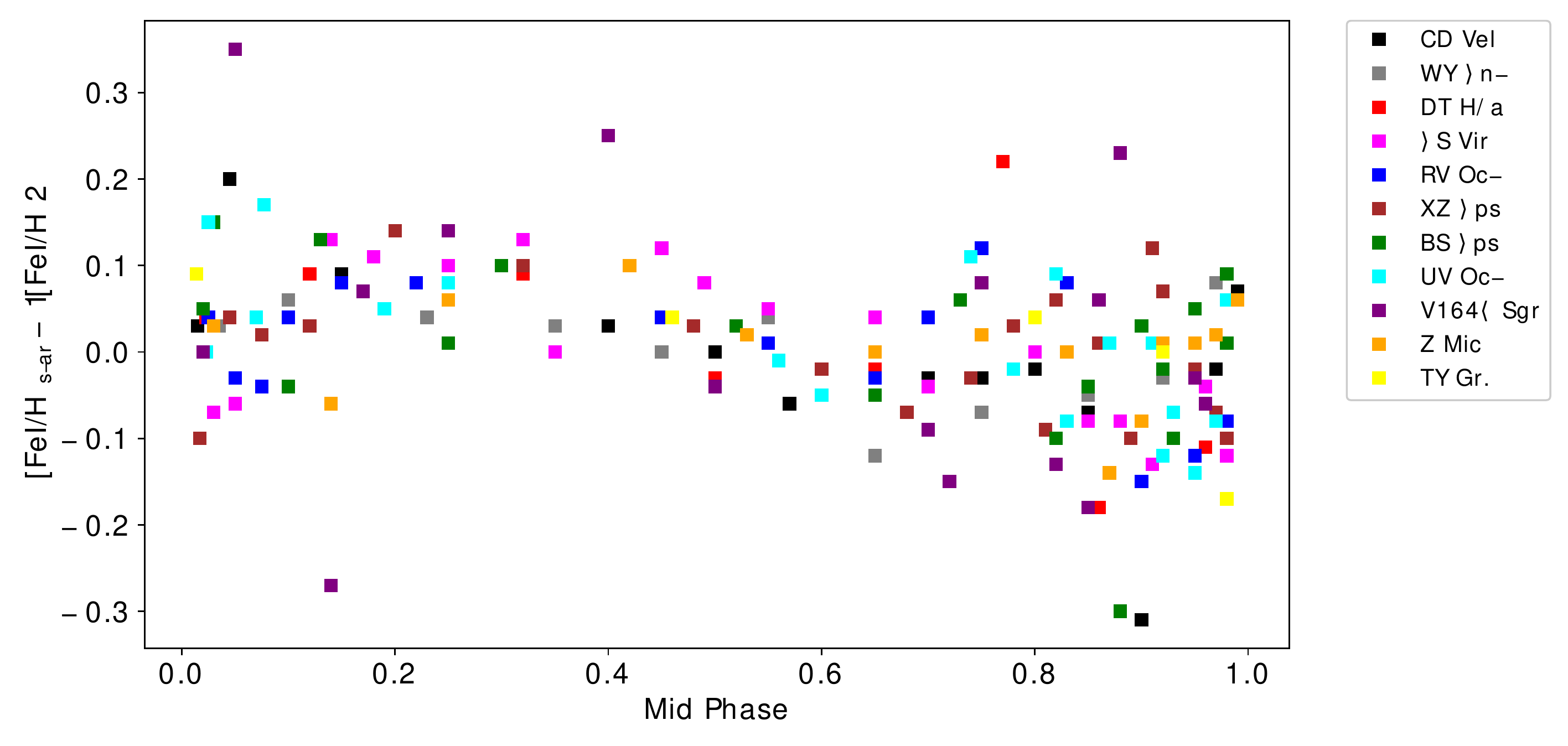}
    \caption{Plot displaying the Mid Phase in the x-axis and the $\mathrm{[Fe I/H]_{Star}}-\langle \mathrm{[Fe I/H]_{Star}} \rangle$ in the y-axis for the 11 different RR Lyrae stars from \citet{2011ApJS..197...29F}.}
    \label{fig:plotForetal2011}
\end{figure}

In order to further discuss the impact of variability on spectroscopic [Fe/H] determinations, it is worth commenting here the paper by \citet{2011ApJS..197...29F}, where the abundances of 11 RR Lyrae ab-type variables were studied. They analyzed more than 2300 high resolution spectra from the 2.5m du Pont telescope distributed along the period curve of each variable. In Figure~\ref{fig:plotForetal2011} we overplot the [Fe/H] abundances of each variable, where each single measurement of each variable was divided by the corresponding mean [Fe/H] abundance. Figure~\ref{fig:plotForetal2011} indicates that a variable can show a variation in [Fe I/H] up to $\pm0.2$ dex, depending on the phase. This means that a variable can have an excursion up to $0.4$ dex in its [Fe/H] content. In addition the mean [Fe I/H] behavior is quite regular with the phase, at least between Mid Phase $0$ and $0.6$. At Mid Phase$=0$ the deviation from the mean iron abundance is around zero. Then the deviation increases toward higher metallicities reaching its maximum at Mid Phase$=0.2$. It remains constant until Mid Phase$=0.4$ and then returns to zero at a Mid Phase$=0.6$. After {that} the behaviors is quite chaotic with some variables showing a positive deviation of $0.1$ dex or more, while others show a negative deviation of $-0.15$ dex or more. On the other hand the maximum deviation in this plot is between $-0.3$ and $+0.3$ dex. Such finding is important since it is an independent confirmation of the results obtained in this work. It shows that in these kind of stars the $\text{[Fe I/H]}$ determination might be affected by variability. Furthermore, it provides us with the possibility that the physical reason for the variability of our stars is some kind of pulsations.

Because of that we suggest that, if the variability is due to pulsations, {the [Fe/H] spread difference between variable and non-variable stars} could originate {from} the fact that the standard technique for chemical analysis uses hydrostatic LTE atmospheric models, while pulsating stars have by definition non-hydrostatic atmospheres. Hydrostatic LTE models work because they resemble real atmospheres where, in most of the cases, the gas is static on the large scale and the gas dense enough to allow the light coming from the deepest part of the photosphere to be fully absorbed and re-processed by the different atmospheric layers before leaving the stellar surface. This means that each layer behaves like a black body and for this reason it can be treated by the LTE approximation. This is not strictly true for the spectral lines formed at the top of the photosphere, but it works for most of the lines that are formed in the deeper parts of the atmosphere. In pulsating stars the atmosphere is expanding and contracting radially or tangentially, and during the expansion the gas become less dense and so its behavior deviates from the LTE approximation. Also the pulsation itself could be have a direct impact on the spectroscopic [Fe/H] determination {in the sense that it could alter the atmospheric structure and specifically how temperature and pressure vary with optical depth. Since these variables are assumed to vary like it is predicted in the atmospheric model, a mismatch between the predicted and the real atmospheric structure can easily lead to an underestimation or overestimation of the iron content.}
%%%%%%%%%%%%%%%%%%%%%%%%%%%%%%%%%%%%%%%%%%%%%%%%%%
%%%%%%%%%%%%%%%%%%%%%%%%%%%%%%%%%%%%%%%%%%%%%%%%%%
\section{Summary and Conclusions}\label{sec:conclus}
We present {in this paper} the results of the variability analysis for 17 giant stars in the GC NGC~3201 that were studied {previously} to obtain spectroscopic metallicity $\text{[Fe/H]}$. {The aim of our study is to check} if variability {affects} somehow the {determination of the} iron content of {RGB and AGB} stars {in globular clusters} as obtained from the spectra using standard techniques. We performed the Generalized Lomb-Scargle and the Phase Dispersion Minimization methods along with  a set of discarding criteria in order to check for variables. Of the total sample of 17 giant stars, 8 targets were classified  as variable stars, 2 as possible variables, and 7 as non-variable.

After obtaining the main parameters (period $P$ and amplitude $A$) of the 8 variable stars for both $V$ and $I$ filters, we revised the results from \citet{2015ApJ...801...69M} regarding their iron abundance. We also analyzed our variable stars according to their period, amplitude, position in the CMD and the shape of the light curve, comparing them with known variables, concluding that most probable reason for the variability is radial or tangential pulsation. We revised also the results from \citet{2013ApJ...764L...7S} to see if there is some difference depending on the different methods used to obtain the metallicity. We conclude that \textit{stars having a larger spread in {their} [Fe/H] \textbf{{estimates}} are variable, while non-variable have an \textbf{{observed}} iron content more concentrated around the mean value of the cluster.} This means that variable stars can display extreme values of metallicity (both toward metal-richer and metal-poorer ends), while stars that do not present variability display values of their iron abundance closer to the mean abundance of the total sample. Furthermore, this finding is consistent with the metal-poor variable RGB star found in \citet{2018A&A...620A..96M}. More data for both photometric and spectroscopic analysis are needed in order to further confirm our results, but nevertheless this is a step forward to determine the nature of the variable star and its influence on metallicity. Finally, our result suggests that the spectroscopic determination of the iron abundance of giant stars in NGC~3201 is affected in some way by the variability of such stars and point toward the fact that this cluster does not have an intrinsic iron spread.
%%%%%%%%%%%%%%%%%%%%%%%%%%%%%%%%%%%%%%%%%%%%%%%%%%

\acknowledgments
SV gratefully acknowledges the support provided by Fondecyt reg. n. 1170518. This work is based on data acquired at Complejo Astron\'{o}mico El Leoncito, operated under agreement between the Consejo Nacional de Investigaciones Cient\'{\i}ficas y T\'{e}cnicas de la Rep\'{u}blica Argentina and the National Universities of La Plata, C\'{o}rdoba and San Juan. Furthermore, this work made use of the PyAstronomy package collection and we are thankful at the PyA group for answering our doubts regarding their packages.
Finally, the authors gratefully acknowledge the comments and suggestions of the anonymous referee
that greatly improved and strengthen this paper.

%\bibliography{Manuscript}

\begin{thebibliography}{}
\expandafter\ifx\csname natexlab\endcsname\relax\def\natexlab#1{#1}\fi
\providecommand{\url}[1]{\href{#1}{#1}}
\providecommand{\dodoi}[1]{doi:~\href{http://doi.org/#1}{\nolinkurl{#1}}}
\providecommand{\doeprint}[1]{\href{http://ascl.net/#1}{\nolinkurl{http://ascl.net/#1}}}
\providecommand{\doarXiv}[1]{\href{https://arxiv.org/abs/#1}{\nolinkurl{https://arxiv.org/abs/#1}}}

\bibitem[{{Arellano Ferro} {et~al.}(2014){Arellano Ferro}, {Ahumada},
  {Calder{\'o}n}, \& {Kains}}]{2014RMxAA..50..307A}
{Arellano Ferro}, A., {Ahumada}, J.~A., {Calder{\'o}n}, J.~H., \& {Kains}, N.
  2014, \rmxaa, 50, 307.
\newblock \doarXiv{1406.2635}

\bibitem[{{Bellazzini} {et~al.}(2008){Bellazzini}, {Ibata}, {Chapman},
  {Mackey}, {Monaco}, {Irwin}, {Martin}, {Lewis}, \&
  {Dalessandro}}]{2008AJ....136.1147B}
{Bellazzini}, M., {Ibata}, R.~A., {Chapman}, S.~C., {et~al.} 2008, \aj, 136,
  1147, \dodoi{10.1088/0004-6256/136/3/1147}

\bibitem[{{Bergemann} {et~al.}(2012){Bergemann}, {Lind}, {Collet}, {Magic}, \&
  {Asplund}}]{2012MNRAS.427...27B}
{Bergemann}, M., {Lind}, K., {Collet}, R., {Magic}, Z., \& {Asplund}, M. 2012,
  \mnras, 427, 27, \dodoi{10.1111/j.1365-2966.2012.21687.x}

\bibitem[{{Carretta} {et~al.}(2009{\natexlab{a}}){Carretta}, {Bragaglia},
  {Gratton}, \& {Lucatello}}]{2009A&A...505..139C}
{Carretta}, E., {Bragaglia}, A., {Gratton}, R., \& {Lucatello}, S.
  2009{\natexlab{a}}, \aap, 505, 139, \dodoi{10.1051/0004-6361/200912097}

\bibitem[{{Carretta} {et~al.}(2009{\natexlab{b}}){Carretta}, {Bragaglia},
  {Gratton}, {Lucatello}, {Catanzaro}, {Leone}, {Bellazzini}, {Claudi},
  {D'Orazi}, {Momany}, {Ortolani}, {Pancino}, {Piotto}, {Recio-Blanco}, \&
  {Sabbi}}]{2009A&A...505..117C}
{Carretta}, E., {Bragaglia}, A., {Gratton}, R.~G., {et~al.} 2009{\natexlab{b}},
  \aap, 505, 117, \dodoi{10.1051/0004-6361/200912096}

\bibitem[{{Carretta} {et~al.}(2010){Carretta}, {Bragaglia}, {Gratton},
  {Lucatello}, {Bellazzini}, {Catanzaro}, {Leone}, {Momany}, {Piotto}, \&
  {D'Orazi}}]{2010A&A...520A..95C}
---. 2010, \aap, 520, A95, \dodoi{10.1051/0004-6361/201014924}

\bibitem[{{Catelan} \& {Smith}(2015)}]{2015pust.book.....C}
{Catelan}, M., \& {Smith}, H.~A. 2015, {Pulsating Stars}

\bibitem[{{Clement} {et~al.}(2001){Clement}, {Muzzin}, {Dufton}, {Ponnampalam},
  {Wang}, {Burford}, {Richardson}, {Rosebery}, {Rowe}, \&
  {Hogg}}]{2001AJ....122.2587C}
{Clement}, C.~M., {Muzzin}, A., {Dufton}, Q., {et~al.} 2001, \aj, 122, 2587,
  \dodoi{10.1086/323719}

\bibitem[{{Czesla} {et~al.}(2019){Czesla}, {Schr{\"o}ter}, {Schneider},
  {Huber}, {Pfeifer}, {Andreasen}, \& {Zechmeister}}]{2019ascl.soft06010C}
{Czesla}, S., {Schr{\"o}ter}, S., {Schneider}, C.~P., {et~al.} 2019, {PyA:
  Python astronomy-related packages}.
\newblock \doeprint{1906.010}

\bibitem[{{Eggen}(1973)}]{1973ApJ...184..793E}
{Eggen}, O.~J. 1973, \apj, 184, 793, \dodoi{10.1086/152371}

\bibitem[{{Eggen}(1977)}]{1977ApJ...213..767E}
---. 1977, \apj, 213, 767, \dodoi{10.1086/155206}

\bibitem[{{For} {et~al.}(2011){For}, {Sneden}, \&
  {Preston}}]{2011ApJS..197...29F}
{For}, B.-Q., {Sneden}, C., \& {Preston}, G.~W. 2011, \apjs, 197, 29,
  \dodoi{10.1088/0067-0049/197/2/29}

\bibitem[{{Gonzalez} \& {Wallerstein}(1998)}]{1998AJ....116..765G}
{Gonzalez}, G., \& {Wallerstein}, G. 1998, \aj, 116, 765,
  \dodoi{10.1086/300476}

\bibitem[{{Gratton} {et~al.}(2004){Gratton}, {Sneden}, \&
  {Carretta}}]{2004ARA&A..42..385G}
{Gratton}, R., {Sneden}, C., \& {Carretta}, E. 2004, \araa, 42, 385,
  \dodoi{10.1146/annurev.astro.42.053102.133945}

\bibitem[{{Harris}(1996)}]{1996yCat.7195....0H}
{Harris}, W.~E. 1996, VizieR Online Data Catalog, VII/195

\bibitem[{{Johnson} {et~al.}(2015){Johnson}, {Rich}, {Pilachowski}, {Caldwell},
  {Mateo}, {Bailey}, \& {Crane}}]{2015AJ....150...63J}
{Johnson}, C.~I., {Rich}, R.~M., {Pilachowski}, C.~A., {et~al.} 2015, \aj, 150,
  63, \dodoi{10.1088/0004-6256/150/2/63}

\bibitem[{{Lapenna} {et~al.}(2014){Lapenna}, {Mucciarelli}, {Lanzoni},
  {Ferraro}, {Dalessandro}, {Origlia}, \& {Massari}}]{2014ApJ...797..124L}
{Lapenna}, E., {Mucciarelli}, A., {Lanzoni}, B., {et~al.} 2014, \apj, 797, 124,
  \dodoi{10.1088/0004-637X/797/2/124}

\bibitem[{{Layden} \& {Sarajedini}(2003)}]{2003AJ....125..208L}
{Layden}, A.~C., \& {Sarajedini}, A. 2003, \aj, 125, 208,
  \dodoi{10.1086/344948}

\bibitem[{{Lind} {et~al.}(2012){Lind}, {Bergemann}, \&
  {Asplund}}]{2012MNRAS.427...50L}
{Lind}, K., {Bergemann}, M., \& {Asplund}, M. 2012, \mnras, 427, 50,
  \dodoi{10.1111/j.1365-2966.2012.21686.x}

\bibitem[{{Marino} {et~al.}(2009){Marino}, {Milone}, {Piotto}, {Villanova},
  {Bedin}, {Bellini}, \& {Renzini}}]{2009A&A...505.1099M}
{Marino}, A.~F., {Milone}, A.~P., {Piotto}, G., {et~al.} 2009, \aap, 505, 1099,
  \dodoi{10.1051/0004-6361/200911827}

\bibitem[{{Marino} {et~al.}(2011){Marino}, {Milone}, {Piotto}, {Villanova},
  {Gratton}, {D'Antona}, {Anderson}, {Bedin}, {Bellini}, {Cassisi}, {Geisler},
  {Renzini}, \& {Zoccali}}]{2011ApJ...731...64M}
---. 2011, \apj, 731, 64, \dodoi{10.1088/0004-637X/731/1/64}

\bibitem[{{Marino} {et~al.}(2015){Marino}, {Milone}, {Karakas}, {Casagrand e},
  {Yong}, {Shingles}, {Da Costa}, {Norris}, {Stetson}, {Lind}, {Asplund},
  {Collet}, {Jerjen}, {Sbordone}, {Aparicio}, \&
  {Cassisi}}]{2015MNRAS.450..815M}
{Marino}, A.~F., {Milone}, A.~P., {Karakas}, A.~I., {et~al.} 2015, \mnras, 450,
  815, \dodoi{10.1093/mnras/stv420}

\bibitem[{{Minniti} {et~al.}(2010){Minniti}, {Lucas}, {Emerson}, {Saito},
  {Hempel}, {Pietrukowicz}, {Ahumada}, {Alonso}, {Alonso-Garcia}, {Arias},
  {Bandyopadhyay}, {Barb{\'a}}, {Barbuy}, {Bedin}, {Bica}, {Borissova},
  {Bronfman}, {Carraro}, {Catelan}, {Clari{\'a}}, {Cross}, {de Grijs},
  {D{\'e}k{\'a}ny}, {Drew}, {Fari{\~n}a}, {Feinstein}, {Fern{\'a}ndez
  Laj{\'u}s}, {Gamen}, {Geisler}, {Gieren}, {Goldman}, {Gonzalez}, {Gunthardt},
  {Gurovich}, {Hambly}, {Irwin}, {Ivanov}, {Jord{\'a}n}, {Kerins}, {Kinemuchi},
  {Kurtev}, {L{\'o}pez-Corredoira}, {Maccarone}, {Masetti}, {Merlo},
  {Messineo}, {Mirabel}, {Monaco}, {Morelli}, {Padilla}, {Palma}, {Parisi},
  {Pignata}, {Rejkuba}, {Roman-Lopes}, {Sale}, {Schreiber}, {Schr{\"o}der},
  {Smith}, {}, {Soto}, {Tamura}, {Tappert}, {Thompson}, {Toledo}, {Zoccali}, \&
  {Pietrzynski}}]{2010NewA...15..433M}
{Minniti}, D., {Lucas}, P.~W., {Emerson}, J.~P., {et~al.} 2010, \na, 15, 433,
  \dodoi{10.1016/j.newast.2009.12.002}

\bibitem[{{Mu{\~n}oz} {et~al.}(2013){Mu{\~n}oz}, {Geisler}, \&
  {Villanova}}]{2013MNRAS.433.2006M}
{Mu{\~n}oz}, C., {Geisler}, D., \& {Villanova}, S. 2013, \mnras, 433, 2006,
  \dodoi{10.1093/mnras/stt868}

\bibitem[{{Mu{\~n}oz} {et~al.}(2018){Mu{\~n}oz}, {Geisler}, {Villanova},
  {Saviane}, {Cort{\'e}s}, {Dias}, {Cohen}, {Mauro}, \& {Moni
  Bidin}}]{2018A&A...620A..96M}
{Mu{\~n}oz}, C., {Geisler}, D., {Villanova}, S., {et~al.} 2018, \aap, 620, A96,
  \dodoi{10.1051/0004-6361/201833373}

\bibitem[{{Mucciarelli} {et~al.}(2017){Mucciarelli}, {Bellazzini}, {Ibata},
  {Romano}, {Chapman}, \& {Monaco}}]{2017A&A...605A..46M}
{Mucciarelli}, A., {Bellazzini}, M., {Ibata}, R., {et~al.} 2017, \aap, 605,
  A46, \dodoi{10.1051/0004-6361/201730707}

\bibitem[{{Mucciarelli} {et~al.}(2015){Mucciarelli}, {Lapenna}, {Massari},
  {Ferraro}, \& {Lanzoni}}]{2015ApJ...801...69M}
{Mucciarelli}, A., {Lapenna}, E., {Massari}, D., {Ferraro}, F.~R., \&
  {Lanzoni}, B. 2015, \apj, 801, 69, \dodoi{10.1088/0004-637X/801/1/69}

\bibitem[{{Origlia} {et~al.}(2003){Origlia}, {Ferraro}, {Bellazzini}, \&
  {Pancino}}]{2003ApJ...591..916O}
{Origlia}, L., {Ferraro}, F.~R., {Bellazzini}, M., \& {Pancino}, E. 2003, \apj,
  591, 916, \dodoi{10.1086/375363}

\bibitem[{{Origlia} {et~al.}(2013){Origlia}, {Massari}, {Rich}, {Mucciarelli},
  {Ferraro}, {Dalessandro}, \& {Lanzoni}}]{2013ApJ...779L...5O}
{Origlia}, L., {Massari}, D., {Rich}, R.~M., {et~al.} 2013, \apjl, 779, L5,
  \dodoi{10.1088/2041-8205/779/1/L5}

\bibitem[{{Origlia} {et~al.}(2011){Origlia}, {Rich}, {Ferraro}, {Lanzoni},
  {Bellazzini}, {Dalessandro}, {Mucciarelli}, {Valenti}, \&
  {Beccari}}]{2011ApJ...726L..20O}
{Origlia}, L., {Rich}, R.~M., {Ferraro}, F.~R., {et~al.} 2011, \apjl, 726, L20,
  \dodoi{10.1088/2041-8205/726/2/L20}

\bibitem[{{Pancino} {et~al.}(2011){Pancino}, {Mucciarelli}, {Sbordone},
  {Bellazzini}, {Pasquini}, {Monaco}, \& {Ferraro}}]{2011A&A...527A..18P}
{Pancino}, E., {Mucciarelli}, A., {Sbordone}, L., {et~al.} 2011, \aap, 527,
  A18, \dodoi{10.1051/0004-6361/201016024}

\bibitem[{{Saito} {et~al.}(2012){Saito}, {Hempel}, {Minniti}, {Lucas},
  {Rejkuba}, {Toledo}, {Gonzalez}, {Alonso-Garc{\'\i}a}, {Irwin},
  {Gonzalez-Solares}, {Hodgkin}, {Lewis}, {Cross}, {Ivanov}, {Kerins},
  {Emerson}, {Soto}, {Am{\^o}res}, {Gurovich}, {D{\'e}k{\'a}ny}, {Angeloni},
  {Beamin}, {Catelan}, {Padilla}, {Zoccali}, {Pietrukowicz}, {Moni Bidin},
  {Mauro}, {Geisler}, {Folkes}, {Sale}, {Borissova}, {Kurtev}, {Ahumada},
  {Alonso}, {Adamson}, {Arias}, {Band yopadhyay}, {Barb{\'a}}, {Barbuy},
  {Baume}, {Bedin}, {Bellini}, {Benjamin}, {Bica}, {Bonatto}, {Bronfman},
  {Carraro}, {Chen{\`e}}, {Clari{\'a}}, {Clarke}, {Contreras}, {Corvill{\'o}n},
  {de Grijs}, {Dias}, {Drew}, {Fari{\~n}a}, {Feinstein},
  {Fern{\'a}ndez-Laj{\'u}s}, {Gamen}, {Gieren}, {Goldman},
  {Gonz{\'a}lez-Fern{\'a}ndez}, {Grand }, {Gunthardt}, {Hambly}, {Hanson},
  {He{\l}miniak}, {Hoare}, {Huckvale}, {Jord{\'a}n}, {Kinemuchi}, {Longmore},
  {L{\'o}pez-Corredoira}, {Maccarone}, {Majaess}, {Mart{\'\i}n}, {Masetti},
  {Mennickent}, {Mirabel}, {Monaco}, {Morelli}, {Motta}, {Palma}, {Parisi},
  {Parker}, {Pe{\~n}aloza}, {Pietrzy{\'n}ski}, {Pignata}, {Popescu}, {Read},
  {Rojas}, {Roman-Lopes}, {Ruiz}, {Saviane}, {Schreiber}, {Schr{\"o}der},
  {Sharma}, {Smith}, {Sodr{\'e}}, {Stead}, {Stephens}, {Tamura}, {Tappert},
  {Thompson}, {Valenti}, {Vanzi}, {Walton}, {Weidmann}, \&
  {Zijlstra}}]{2012A&A...537A.107S}
{Saito}, R.~K., {Hempel}, M., {Minniti}, D., {et~al.} 2012, \aap, 537, A107,
  \dodoi{10.1051/0004-6361/201118407}

\bibitem[{{Simmerer} {et~al.}(2013){Simmerer}, {Ivans}, {Filler}, {Francois},
  {Charbonnel}, {Monier}, \& {James}}]{2013ApJ...764L...7S}
{Simmerer}, J., {Ivans}, I.~I., {Filler}, D., {et~al.} 2013, \apjl, 764, L7,
  \dodoi{10.1088/2041-8205/764/1/L7}

\bibitem[{{Stellingwerf}(1978)}]{1978ApJ...224..953S}
{Stellingwerf}, R.~F. 1978, \apj, 224, 953, \dodoi{10.1086/156444}

\bibitem[{{Szymanski}(2005)}]{2005AcA....55...43S}
{Szymanski}, M.~K. 2005, \actaa, 55, 43.
\newblock \doarXiv{astro-ph/0602018}

\bibitem[{{Udalski} {et~al.}(1997){Udalski}, {Kubiak}, \&
  {Szymanski}}]{1997AcA....47..319U}
{Udalski}, A., {Kubiak}, M., \& {Szymanski}, M. 1997, \actaa, 47, 319.
\newblock \doarXiv{astro-ph/9710091}

\bibitem[{{Udalski} {et~al.}(2008){Udalski}, {Szymanski}, {Soszynski}, \&
  {Poleski}}]{2008AcA....58...69U}
{Udalski}, A., {Szymanski}, M.~K., {Soszynski}, I., \& {Poleski}, R. 2008,
  \actaa, 58, 69.
\newblock \doarXiv{0807.3884}

\bibitem[{{Villanova} {et~al.}(2013){Villanova}, {Geisler}, {Carraro}, {Moni
  Bidin}, \& {Mu{\~n}oz}}]{2013ApJ...778..186V}
{Villanova}, S., {Geisler}, D., {Carraro}, G., {Moni Bidin}, C., \&
  {Mu{\~n}oz}, C. 2013, \apj, 778, 186, \dodoi{10.1088/0004-637X/778/2/186}

\bibitem[{{Yepez} {et~al.}(2018){Yepez}, {Arellano Ferro}, {Muneer}, \&
  {Giridhar}}]{2018RMxAA..54...15Y}
{Yepez}, M.~A., {Arellano Ferro}, A., {Muneer}, S., \& {Giridhar}, S. 2018,
  \rmxaa, 54, 15.
\newblock \doarXiv{1709.00044}

\bibitem[{{Yong} {et~al.}(2014){Yong}, {Roederer}, {Grundahl}, {Da Costa},
  {Karakas}, {Norris}, {Aoki}, {Fishlock}, {Marino}, {Milone}, \&
  {Shingles}}]{2014MNRAS.441.3396Y}
{Yong}, D., {Roederer}, I.~U., {Grundahl}, F., {et~al.} 2014, \mnras, 441,
  3396, \dodoi{10.1093/mnras/stu806}

\bibitem[{{Zechmeister} \& {K{\"u}rster}(2009)}]{2009A&A...496..577Z}
{Zechmeister}, M., \& {K{\"u}rster}, M. 2009, \aap, 496, 577,
  \dodoi{10.1051/0004-6361:200811296}

\end{thebibliography}
%\bibliographystyle{aasjournal}

\end{document}